\documentclass[modern]{aastex63}

\usepackage{todonotes}

\received{TBD, 2020}
\submitjournal{ApJ}

\shorttitle{Turbulence in a Radiative Shock Wave}
\shortauthors{Raymond et al.}
\graphicspath{{./}{figures/}}

\begin{document}

\def\arcmin{$^\prime$}
\def\arcsec{\arcmin \arcmin}
\def\kms{$\rm km~s^{-1}$}
\def\cm3{$\rm cm^{-3}$}


\title{Turbulence and Energetic Particles in Radiative Shock Waves in the Cygnus Loop II: Development of Postshock Turbulence}

\correspondingauthor{John C. Raymond}
\email{jraymond@cfa.harvard.edu}

\author[0000-0002-7868-1622]{John C. Raymond}
\affiliation{Center for Astrophysics | Harvard \& Smithsonian \\
60 Garden St. \\
Cambridge, MA 02138, USA}

\author[0000-0002-7597-6935]{Jonathan D. Slavin}
\affil{Center for Astrophysics | Harvard \& Smithsonian, 60 Garden St., Cambridge, MA 02138, USA}

\author[0000-0003-2379-6518]{William P. Blair}
\affiliation{The Henry A. Rowland Department of Physics and Astronomy, 
Johns Hopkins University, 3400 N. Charles Street, Baltimore, MD, 21218, USA; 
wblair@jhu.edu}

\author[0000-0002-7924-3253]{Igor V. Chilingarian}
\affil{Center for Astrophysics | Harvard \& Smithsonian, 60 Garden St., Cambridge, MA 02138, USA}
\affil{Sternberg Astronomical Institute, M.V.Lomonosov Moscow State University, Universitetsky prospect 13, Moscow, 119234, Russia}

\author[0000-0001-5817-5944]{Blakesley Burkhart}
\affil{Center for Computational Astrophysics, Flatiron Institute, 162 Fifth Ave., New York, NY 10010, USA}
\affil{Department of Physics and Astronomy, Rutgers, The State University of New Jersey, 136 Frelinghuysen Rd., Piscataway, NJ 08854, USA}

\author[0000-0001-8858-1943]{Ravi Sankrit}
\affil{Space Telescope Science Institute, Baltimore, MD, USA}




\begin{abstract}
Radiative shock waves in the Cygnus Loop and other supernova remnants show different morphologies in [O III] and H$\alpha$ emission.  We use HST spectra and narrowband images to study the development of turbulence in the cooling region behind a shock on the west limb of the Cygnus Loop.  We refine our earlier estimates of shock parameters that were based upon ground-based spectra, including ram pressure, vorticity and magnetic field strength. We apply several techniques, including Fourier power spectra and the Rolling Hough Transform, to quantify the shape of the rippled shock front as viewed in different emission lines.  We assess the relative importance of thermal instabilities, the thin shell instability, upstream density variations, and upstream magnetic field variations in producing the observed structure. 

\end{abstract}



\keywords{shocks --- supernova remnants --- plasma astrophysics --- interstellar magnetic fields --- turbulence}


\section{Introduction} \label{sec:intro}

\begin{figure}
\center
\includegraphics[width=3in]{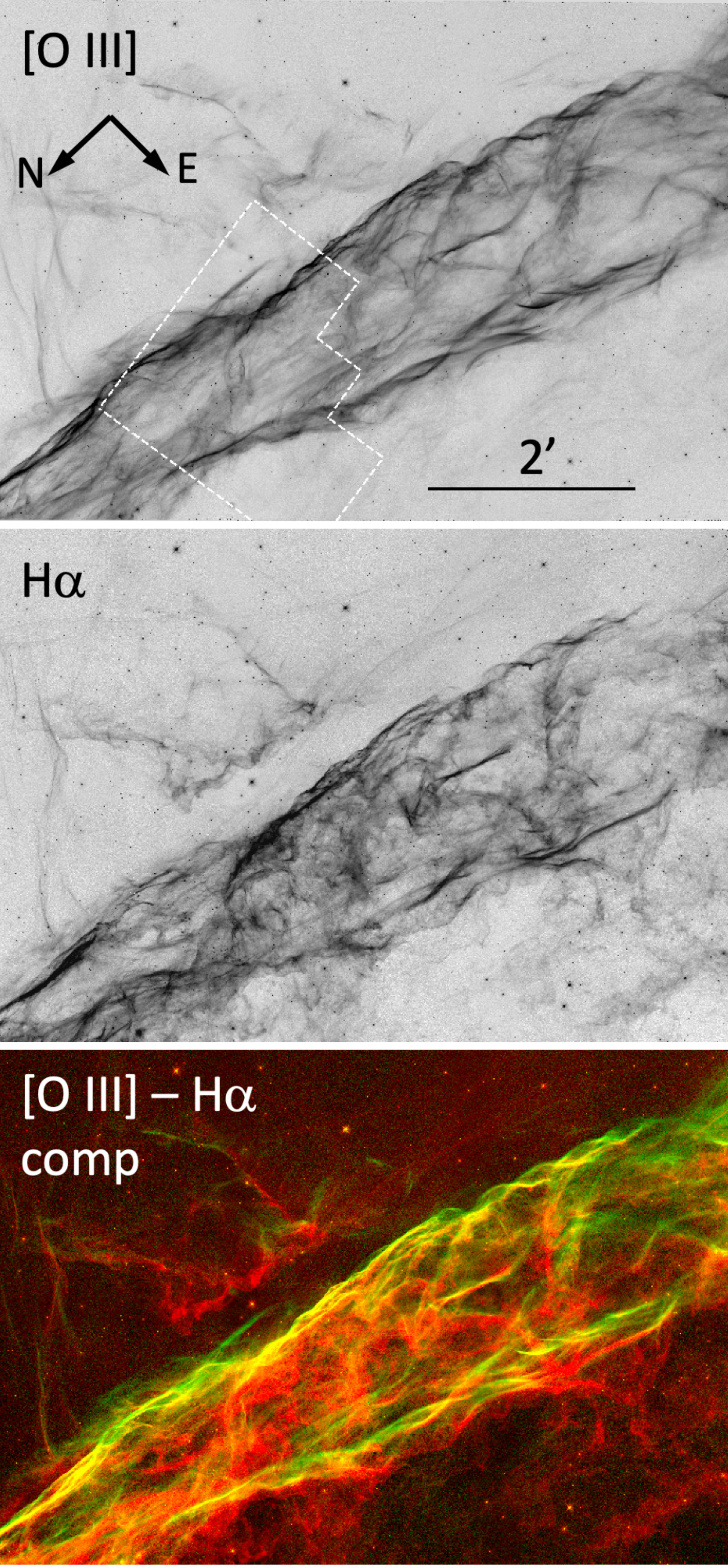}
\caption{The mosaics made from six WFC3 fields in the western Cygnus Loop are shown in [O~III] (top) and H$\alpha$ (middle), demonstrating the dramatic change in morphology.  (Note: the F657N filter used for the Heritage mosaic was broad enough to include some emission from [N~II] as well.) The same two data frames are shown in the bottom panel ([O~III] in green, H$\alpha$ in red) to show the spatial alignments. The dashed white outline in the top panel shows the location of the WFPC2 data from 1997, which also highlights the region of filaments of primary interest in this paper.
\label{heritage}
}
\end{figure}

In 2015, the Hubble Heritage project obtained a set of six-field WFC3 mosaics of narrowband images of a portion of the western Cygnus Loop in emission lines of [O~III], H$\alpha$+[N~II], and [S~II].\footnote{A color version of the resulting combined image was released for HST's 25th anniversary.  This image and the data are available as a High Level Science Product from the MAST archive; see \url{https://archive.stsci.edu/prepds/heritage/veil/}.}  These mosaics show a remarkable morphological difference between the smooth, sinuous filaments seen in [O~III] and clumpy, disordered structure in the [S~II] and H$\alpha$ emission produced at lower temperatures, as shown in Fig-\ref{heritage}.  Comparison of these images with WFPC2 images obtained in 1997 provides accurate proper motions.  
The proper motions translate into shock velocities given the known distance of 735$\pm$25 pc to the Cygnus Loop \citep{fesen18}.


In \citet{raymond20} (hereafter Paper I), we reported ground-based long-slit spectra with the Binospec spectrograph on the MMT telescope cutting across filaments in this region.  In this paper, we build on these earlier results using additional HST WFC3 imagery of a portion of the Heritage mosaic with an expanded set of filters.  We also present long-slit optical and UV spectra from the HST STIS spectrograph.  We strive to understand the striking change in morphology in the cooling gas behind the expanding shock wave and to investigate the development of turbulence in the cooling, recombining flow.

\subsection{Interstellar turbulence}

Turbulence is ubiquitously observed in the interstellar medium (ISM) in the form of density and velocity fluctuations inferred from radio observations \citep{armstrong95} as well as magnetic fluctuations measured directly by the Voyager spacecraft \citep{burlaga18}.  Turbulence in the ISM spans scales from at least kilometers to tens of parsecs and seems to follow a Kolmogorov power law \citep{spangler01, burkhart10, krumholz16}.  The turbulence is generally attributed to supernova explosions, which inject energy at large scales as the shocks slow down to merge with the ISM.  Energy liberated by mass transport in an accretion-like flow in the Galactic disk could also be important \citep{krumholz18}.   Under the interpretation of a Kolmogorov energy cascade from large to small scales, the existence of a single power law over the full range of scales suggests that the large-scale energy injection dominates over intermediate-scale energy sources \citep{chepurnov15,pingel18}. 

On large scales, turbulence plays a vital role in the structure and evolution of the ISM.  Compressive turbulence, coupled with gravity and feedback, in molecular clouds helps control the star formation rate \citep{burkhart18}.  Turbulent deformations of the magnetic field are a step in the amplification of the magnetic field by dynamo action \citep{parker71}.  Turbulence is also vital to the acceleration and propagation of cosmic rays \citep{xulazarian20}.  On small scales, turbulence is generated by plasma instabilities in collisionless shocks.  It determines the shock structure and governs the evolution of the particle velocity distributions and the transfer of energy among different particle species, and it can amplify small-scale magnetic fields and accelerate energetic particles \citep{bykov82, blandfordeichler, bell04, zank15}. 
   
This paper is concerned with turbulence on intermediate scales generated by fluid instabilities in the cooling flows behind radiative shock waves \citep{blair91, blair02}. During the encounter of a shock with a density enhancement, the shock transitions from nonradiative to radiative.   Depending on the effective age of the interaction, the shock can be incomplete (i.e. only a partial cooling zone) to fully radiative in nature (i.e. full cooling and recombination zones).  Thermal instabilities can develop in shocks faster than about 120 \kms, a situation that was discussed by \citet{innes87} and explored further by \citet{gaetz88}, \citet{innes92} and \citet{sutherland03}.  A rippled, pressure-driven shell is subject to a thin shell instability, because the ram pressure of the upstream gas and the gas pressure behind the shell are misaligned \citep{vishniac83}.  Any density inhomogeneities in the upstream gas create velocity shear that generates vorticity \citep{klein94, giacalone07, guo12, fraschetti15}.  For typical ISM and shock parameters,  the flow behind a radiative shock makes a transition from gas pressure-dominated to magnetic pressure-dominated (high $\beta$ to low $\beta$) as the gas cools.  Therefore, any nonuniformity in the upstream B field will cause motions and compression or rarefaction of the gas when it cools \citep{raymondcuriel}.  We will attempt to determine how each of these processes helps create the stucture in the Cygnus Loop.

\subsection{Summary of ground-based results}

In Paper I, we obtained a set of adjacent long slit spectra with the Binospec instrument on the MMT telescope \citep{fabricant19} to determine the basic parameters of the shock, though obviously the parameters vary from place to place.  In the region observed, which is in the northern part of the HST images, the shock speed is about 130 \kms\ based on proper motions and the 735$\pm$25 pc distance to the Cygnus Loop \citep{fesen18}.  The ram pressure was determined from the intensity and Doppler velocity of the [O III] line to be about $4 \times 10^{-9} ~\rm dyn~cm^{-2}$, which implies a preshock density of 6 \cm3 \/ for a 130 \kms\ shock.  Downstream electron densities of 100-250 \cm3\/ were found from the [S II] doublet ratio in the region where the gas is partly recombined, and in combination with the preshock density that gives a compression factor of about 50.  The postshock gas is supported by a combination of thermal pressure and magnetic pressure, and we can equate it to the ram pressure to find a postshock magnetic field of 300 $\mu$G.  Dividing by the compression gives 6 $\mu$G for the perpendicular component of the preshock field.  Gradients along and perpendicular to the slit in the Doppler velocity centroids yielded an estimate for the vorticity of $2 \times 10^{-10}~\rm s^{-1}$.      
 
\subsection{This Paper}

In this paper, we attempt to quantify the turbulence in the cooling flow behind the expanding shock front in a portion of the western Cygnus Loop, determine its origin, and understand its evolution.  Even with the exquisite spatial resolution HST images and the ground-based and HST spectra in hand, it is a challenge to separate the turbulent structure from beautiful, but physically less interesting, projection effects.  We employ a number of techniques, including direct comparison of images in different spectral lines, measurement of velocity variations inferred from proper motions, and measurement of the structure by means of autocorrelations, cross-correlations, Fourier transforms, and a technique called the Rolling Hough Transform (RHT)  \citep{duda72,clark14}.  

This paper also revisits some results from Paper I based on the UV and optical spectra from STIS.  In particular, the UV spectra strengthen our conclusion that 1D shock models reproduce the average observed spectra of the complex 3D structure surprisingly well, and they provide new information about the liberation of C and Si from grains.  The very narrow slit optical spectrum improves our estimate of the ram pressure, and the velocity gradients from proper motions improve our vorticity estimate.   We also estimate the kinetic energy of tubulent motions just behind the shock. 

To quantify the structure, we measure the amplitudes and wavelengths of the ripples, the variance in angle to the average shock direction, and the variation in shock speed.  The simple picture of a 1D shock with modest ripples would predict that H$\alpha$ filaments trail [O III] filaments by 1\arcsec\/ to 3\arcsec\/ and that they are somewhat thicker.  While this expectation is met in some places, it more generally is not.  In section 4 we discuss how thermal instabilities, the thin sheet instability, vorticity due to preshock density inhomogeneities, and compression variations due to inhomogeneities in the upstream magnetic field would affect the morphology, and we draw conclusions about the relative importance of each of these mechanisms.

\section{OBSERVATIONS}

\subsection{HST Images}

A set of specialized narrowband images was obtained with WFC3 under Observing Program 15285 in 2018 July. (The spectra discussed below were obtained under the same program ID approximately one year later.)  The data can be obtained from the MAST archive at DOI \url{10.17909/t9-kstd-dw56}. In addition to full single WFC3 field images in the [O~I] $\lambda$6300, [O~II] $\lambda$3727, and the narrow H$\alpha$ filter (F656N), we obtained images with various Quad filters in the lines of [Ne~IV] $\lambda$2420 and [O~III] $\lambda$4363, as well as the two individual members of the [S~II] doublet at $\lambda\lambda$6717, 6732, respectively.  Each exposure with a Quad filter simultaneously obtains images in three other lines at offset positions in the WFC3 field, some of which are of potential interest. In particular, each of the Quad filter images in one of the [S~II] lines also gave an image of an adjacent region in the other [S~II] line, and the image sequence was designed to give exposures in both lines across the bright region of the shock.  In addition, an exposure in C~II] $\lambda$2325 was obtained during each [Ne~IV] exposure, but offset to the south.  A log of the image observations is given in Table 1, where the primary line of interest for the Quad exposures is listed.

We had planned to extract maps of the density from the ratio of the [S~II] lines from the Quad filters, along with maps of the temperature of the [O~III] emitting gas from the ratio of the 4363 \AA\/ line to the 5007 \AA\/ line.  A few of the brighter filaments in the field yielded densities similar to those obtained from the Binospec spectra (Paper I) and [O~III] temperatures similar to those seen in spectra of other radiative shock waves.  However, even after fairly heavy binning, so many of the derived [S~II] ratios lay outside the theoretically allowed range between the high and low density limits that we were unable to obtain a useful map.  Similarly, only a small region yielded potentially useful [O~III] temperatures, and we do not discuss these ratio maps futher.  During each STIS observation (discussed below), we obtained an WFC3 parallel observation on a region to the south of our target region.  Those WFC3 images overlap with ACS images obtained by the Heritage program.  Analysis of those images will be deferred to a future publication.

\begin{table}
\begin{center}
\caption {WFC3 Images}
\begin{tabular}{l l l c }
\hline
\hline
Image   &  Filter$^a$  &  Ion & Exposure \\
\hline
0501010  &  F631N  &  [O I] 6300  &  2660 \\
0501020  &  F373N  &  [O II] 3727  &  2660 \\
0501030  &  F656N  &  H$\alpha$ &  2660 \\
0502010  &  FQ672N  &  [S II]  &  1246 \\
0502020  &  FQ674N  &  [S II] &  1246 \\
05a2010  &  FQ672N  &  [S II] &  1246 \\
05a2020  &  FQ674N  &  [S II] &  1246 \\
0505010  &  FQ437N  &  [O III] 4363 & 2628 \\
0505020  &  FQ437N  &  [O III] 4363 & 2660 \\
0506010  &  FQ243N  &  [Ne IV] 2420 & 2628 \\
0506020  &  FQ243N  &  [Ne IV] 2420 & 2739 \\
Heritage &  F502N   &  [O III] 5007 & 2850  \\
Heritage &  F657N  &  H$\alpha$+[N II] & 1200\\
Heritage &  F673N  & [S II] & 1995 \\
\hline
\end{tabular}
\tablenotetext{a}{The WFC3 Quad filters (FQ) cover roughly \\ 25\% of the normal WFC3 field of view.}
\end{center}
\label{tbl:imagelog}
\end{table}

\begin{figure}
\center
\includegraphics[width=5in]{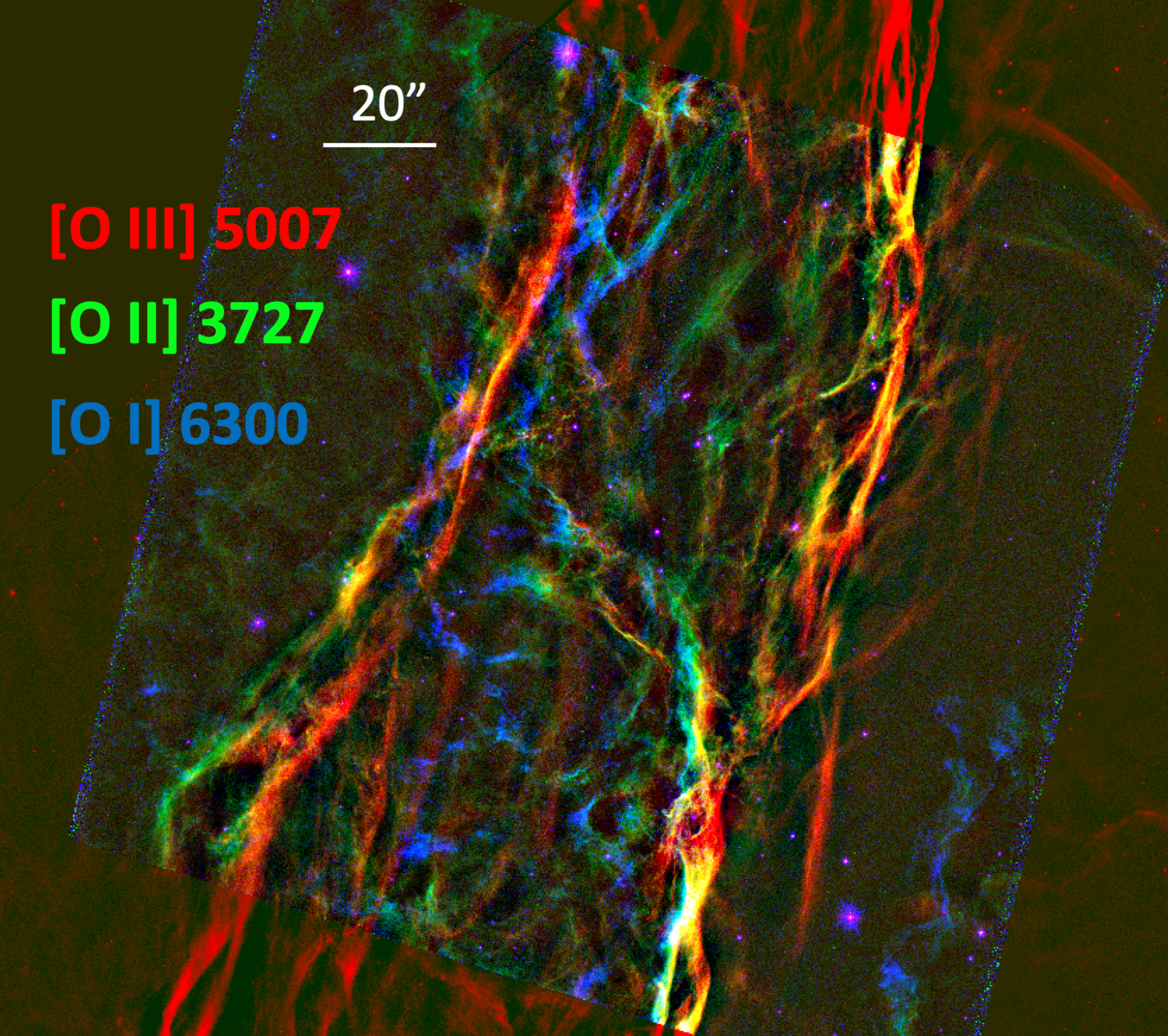}
\caption{An overview showing the gross ionization structure across the observed flamentary structure, where [O~III] is in red, [O~II] in green, and [O~I] blue.  The data frames have been aligned based on stars in the GAIA catalog.  The [O~I] and [O~II] images are from the current program, while the [O~III] image is the mosaic from the Hubble Heritage project.  Yellow regions are bright in both [O~II] and [O~III].  Note how the [O~I] emission seems largely decoupled from the higher ionization lines, although there are places where [O~II] transitions to [O~I].  Figure~\ref{ion-3panel} shows enlargements of sections of this figure to show small scale structure. The tilted inner square is the 162\arcsec\ WFC3 field of view.  North is at the top and east to the left.
\label{overlay}
}
\end{figure}

\begin{figure}
\center
\includegraphics[width=6in]{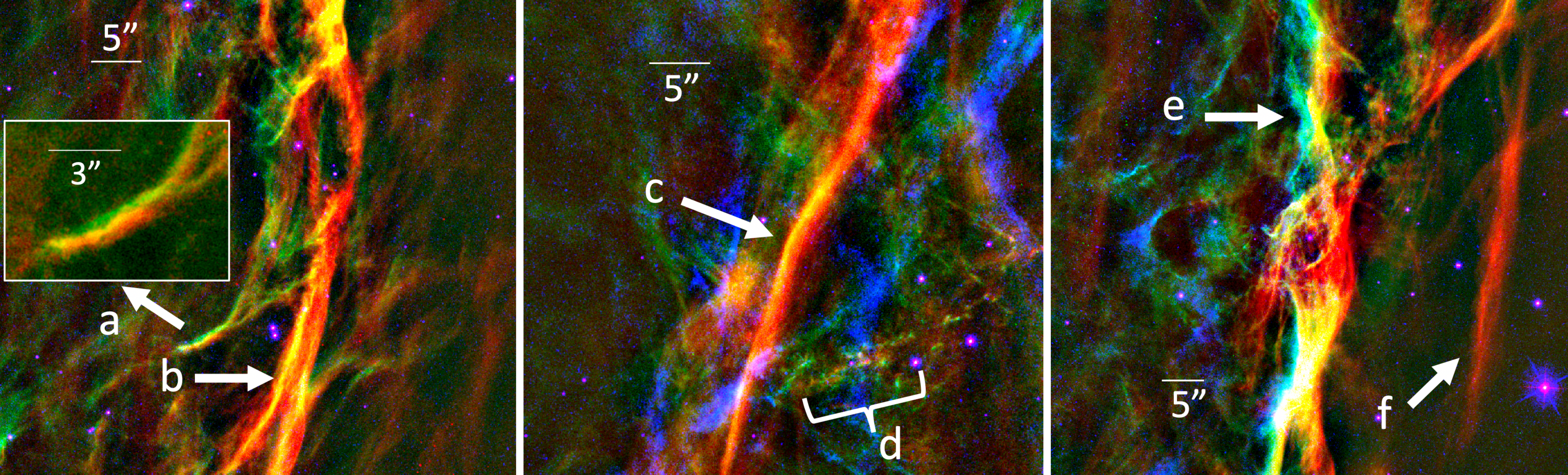}
\caption{Three subregions from the previous Figure are presented to show examples of the complicated fine scale ionization structure. Labeled arrows indicate features described in the text. The inset labeled `a' in the left panel shows one of the cleanest examples of transition from [O~III] to [O~II].  Arrows `b' and `c' indicate narrow filaments where a leading edge in [O~III] is followed by the offset peak in [O~II] which shows as yellow.  Arrow `e' indicates a more extended filament where the [O~II] emission extends further behind the [O~III]. The region denoted by `d' shows a grouping of knotty radiative filaments. Arrow `f' marks a leading edge incomplete shock filament that only appears in [O~III]. As with Figure~\ref{overlay}, north is at the top and east to the left and scale bars are shown.
\label{ion-3panel}
}
\end{figure}

The astrometry provided in the headers of the WFC3 images was not exact.  Therefore, we used a {\sc python} package developed by K.~Grishin to correct the astrometry in HST images.  This package first detected sources, then matched them against the Gaia DR2 catalog and recomputed the projection matrix using a nonlinear $\chi^2$ fitting technique.  Finally, it updated World Coordinate System keywords in the FITS headers. The resulting astrometric accuracy was of order 0\farcs008--0\farcs015. Then, we reprojected all updated HST images from different epochs using SWarp \citep{bertin02} to the same center position and resampled them to 0\farcs04 pixels.  We used the same procedure for the Heritage images and for the WFPC2 images from 1997 that were used to measure proper motions, resampling from the original 0\farcs1 pixels to 0\farcs04.

Figure~\ref{overlay} shows an ionization structure overview, created using the aligned data.  The [O~I] and [O~II] data are from the current program, and the [O~III] image is from the Hubble Heritage program.  Since these are all ions of oxygen, the comparison shows ionization changes in the filaments.  The difference in morphology among the lines is readily apparent.  The [O~III] emission appears as smooth, gently curved filaments, exactly as expected for a thin, slightly rippled sheet of gas seen nearly edge-on, as seen in a wider view in Fig. \ref{heritage}. The [O~II] emission is similar in places to the [O~III], but shows other regions that are more diffuse or even knotty in appearance.  In some cases a filament changes from [O~III] at one end to [O~II] at the other, showing a change in ionization with position.  The [O~I] emission is significantly more clumpy and is largely decoupled from the higher-ionization lines, although there are places where frothy, [O~II] emission transitions into [O~I].  Of course, significant projection effects are also at play in this complicated field. 

In Figure~\ref{ion-3panel}, we show three subsections from Figure~\ref{overlay} to show details. The labeled regions indicate features that will be discussed further below.  Here we give a brief description of the main features seen, which primarily involve the [O~III] and [O~II] emission, which show some correlations.  As the color scale shows, filaments showing as red are dominated by [O~III], and those shown in green are dominated by [O~II].  Hence, a filament like that indicated by arrow `f' is a shock with a very incomplete cooling zone, only showing [O~III] emission.  Filaments such as those marked with `b' and `c' show smooth [O~III] filaments with trailing edges that get bright in [O~II], as expected from the 1D models.  The overlap regions appear yellow.  The filament enlarged in the inset in the left panel shows one of the few filaments where this structure is more resolved, where the filament transitions from red to yellow to green over about 1\arcsec. The region marked with arrow `e' shows a similar offset of [O~II] extending behind the filament, but the interpretation is complicated by the more extended structure of the filaments and likely projection effects. Finally, the region marked ~d' indicates some knotty radiative filaments, many of which have cooled to the point where little [O~III] emission is indicated but [O~II] is strong.

\begin{figure*}
\center
\includegraphics[trim={6.5cm 0 4.5cm 1cm},width=0.49\hsize]{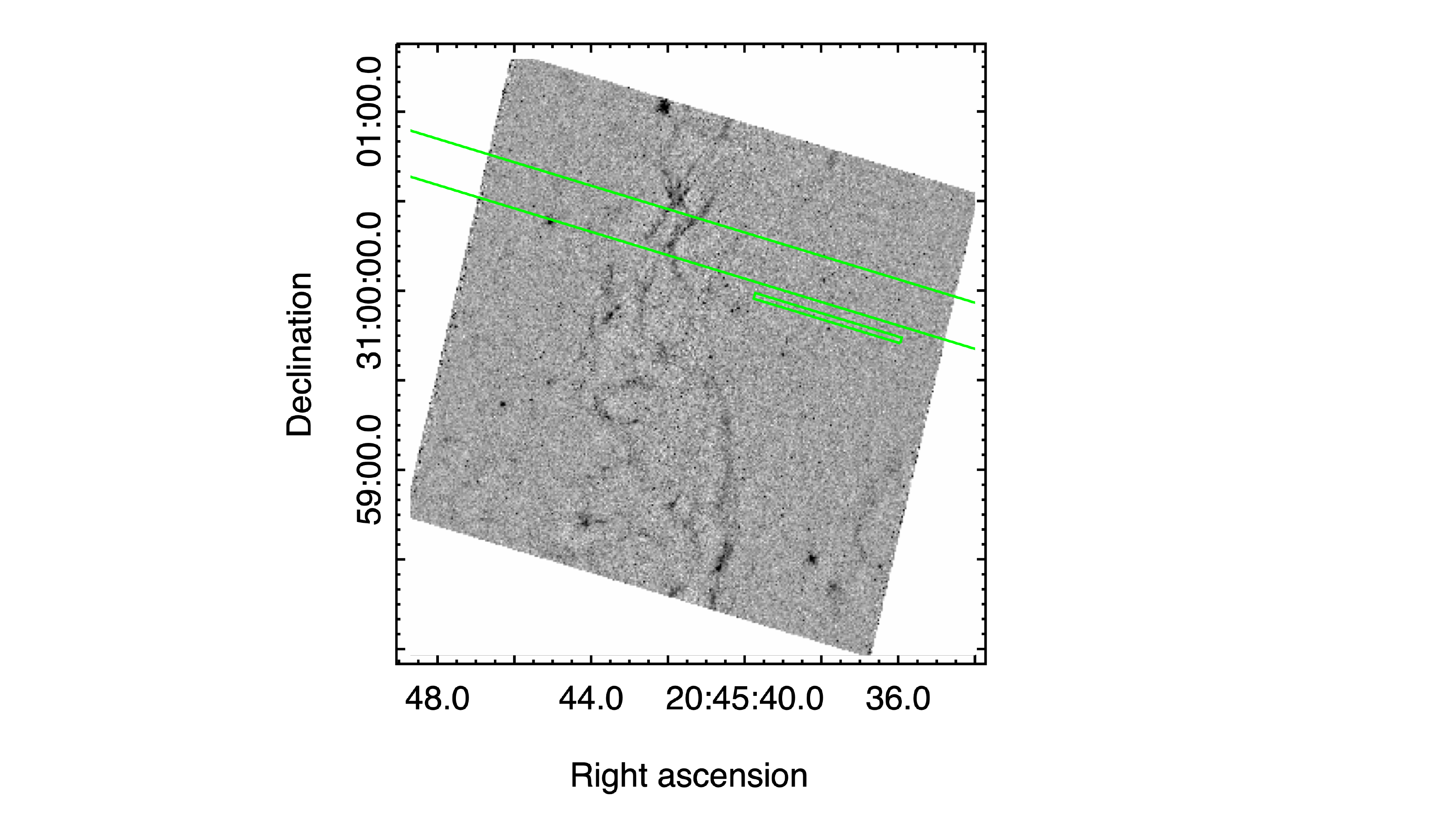}
\includegraphics[trim={6.5cm 0 4.5cm 1cm},width=0.49\hsize]{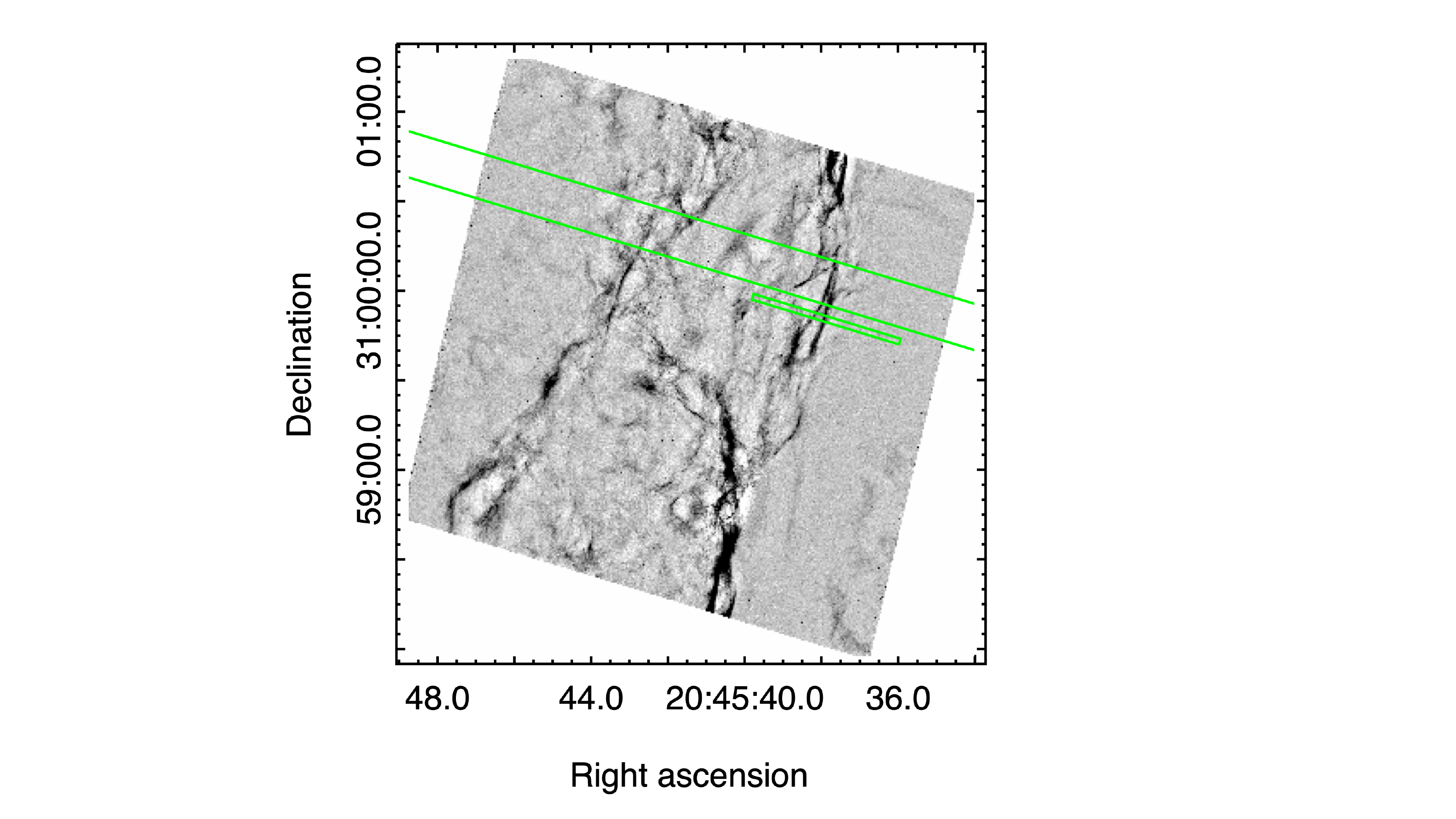}
\caption{WFC3 images in the [O~I] (left) and [O~II] lines (right).  Bright emission regions are shown as black in this figure.  The narrow green rectangle indicates the position of the STIS slit.  The longer green lines indicate the region observed with Binospec in Paper I.
\label{o1_o2}
}
\end{figure*}

\begin{figure*}
\center
\includegraphics[trim={6.5cm 0 4.5cm 1cm},width=0.49\hsize]{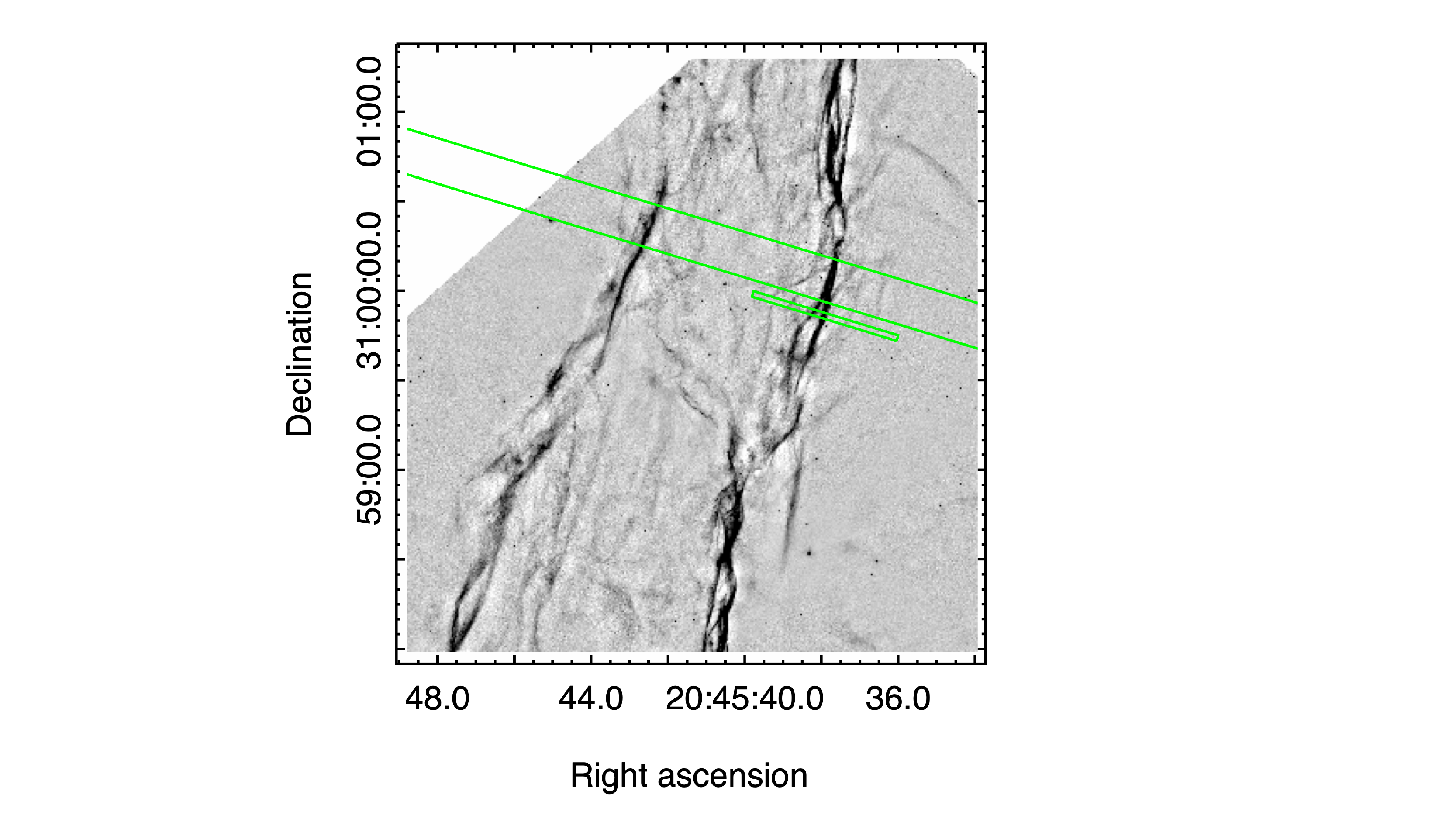}
\includegraphics[trim={6.5cm 0 4.5cm 1cm},width=0.49\hsize]{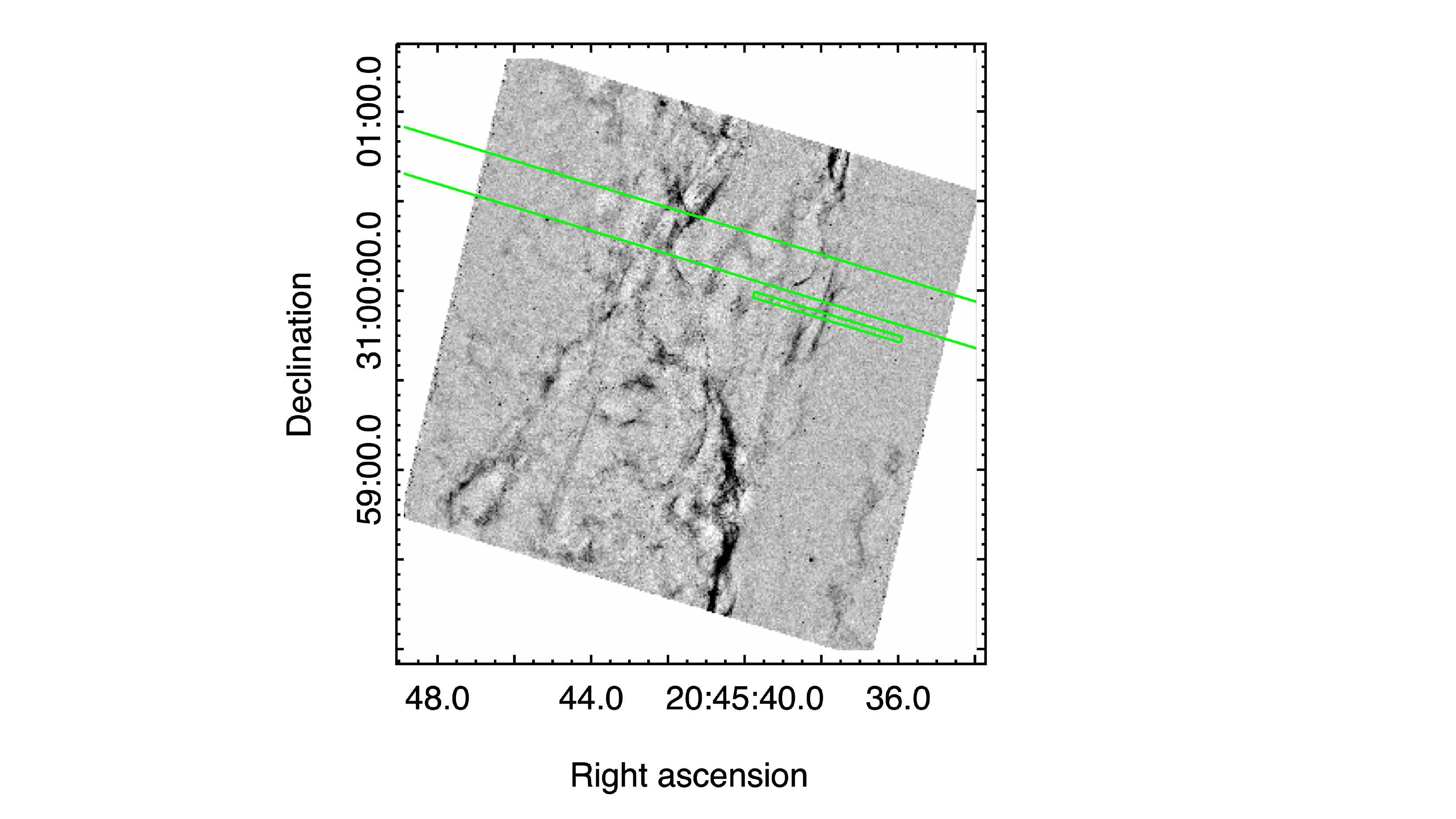}
\caption{Image in [O~III] from the Hubble Heritage mosaic (left) and WFC3 image in H$\alpha$ from the current program (right).  Dark regions in the figure correspond to bright emission.  The narrow green rectangle shows the position of the STIS slit.  The longer green lines indicate the region observed with Binospec in Paper I. 
\label{o3_ha}
}
\end{figure*}

Figures~\ref{o1_o2} and \ref{o3_ha} show the separate images in the individual lines.  The [O~I] emission is relatively faint, but it is generally similar to the H$\alpha$ image.  The [O~II] emission has many features in common with the [O~III] image, but it bears more resemblance to H$\alpha$. In some places the [O~II] seems to trail the [O~III] as would be expected from the 1D models, but in many places a filament shows up in only one line or the other.


\begin{figure*}
  \center
    \includegraphics[width=0.49\hsize]{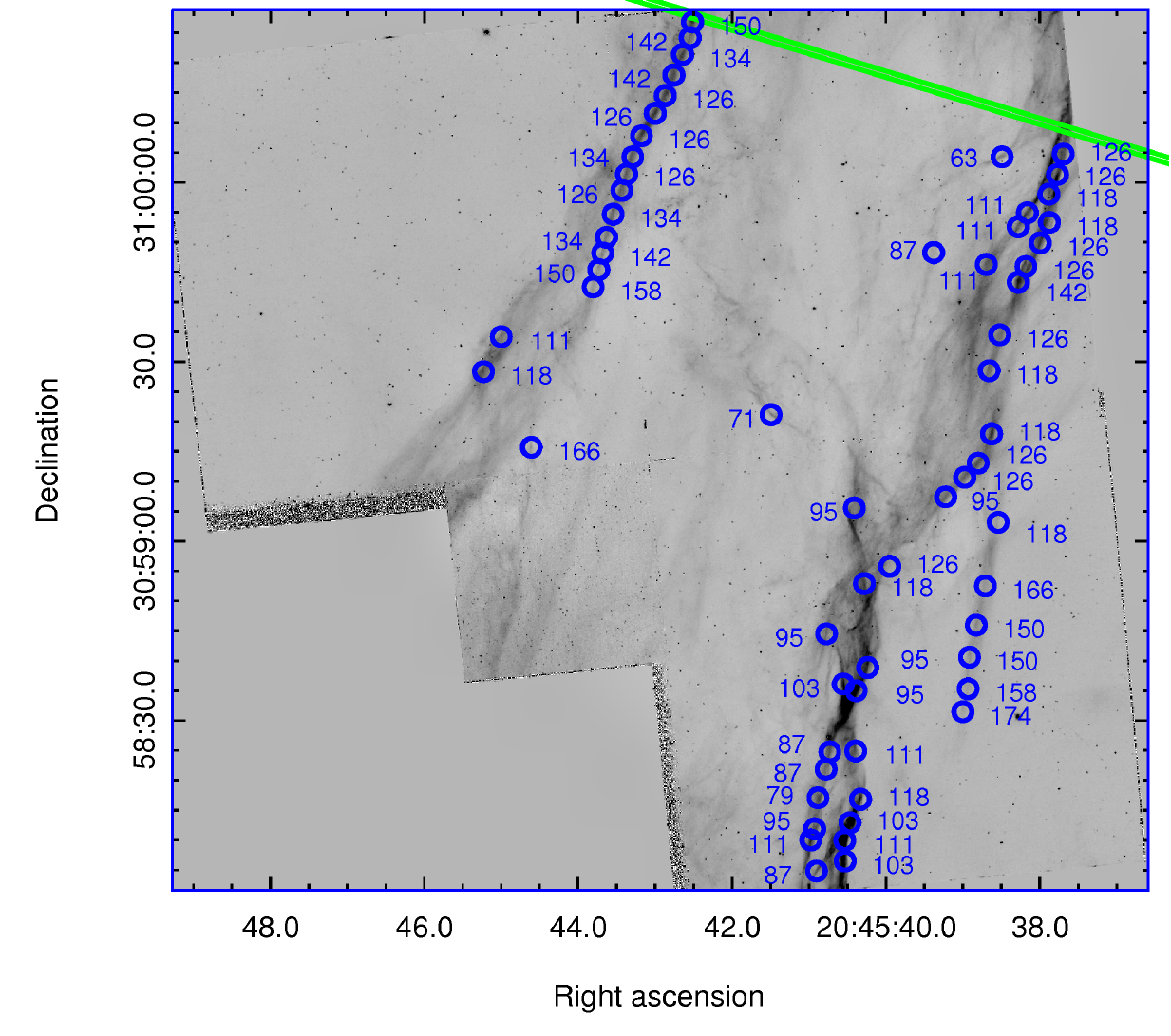}
    \includegraphics[width=0.49\hsize]{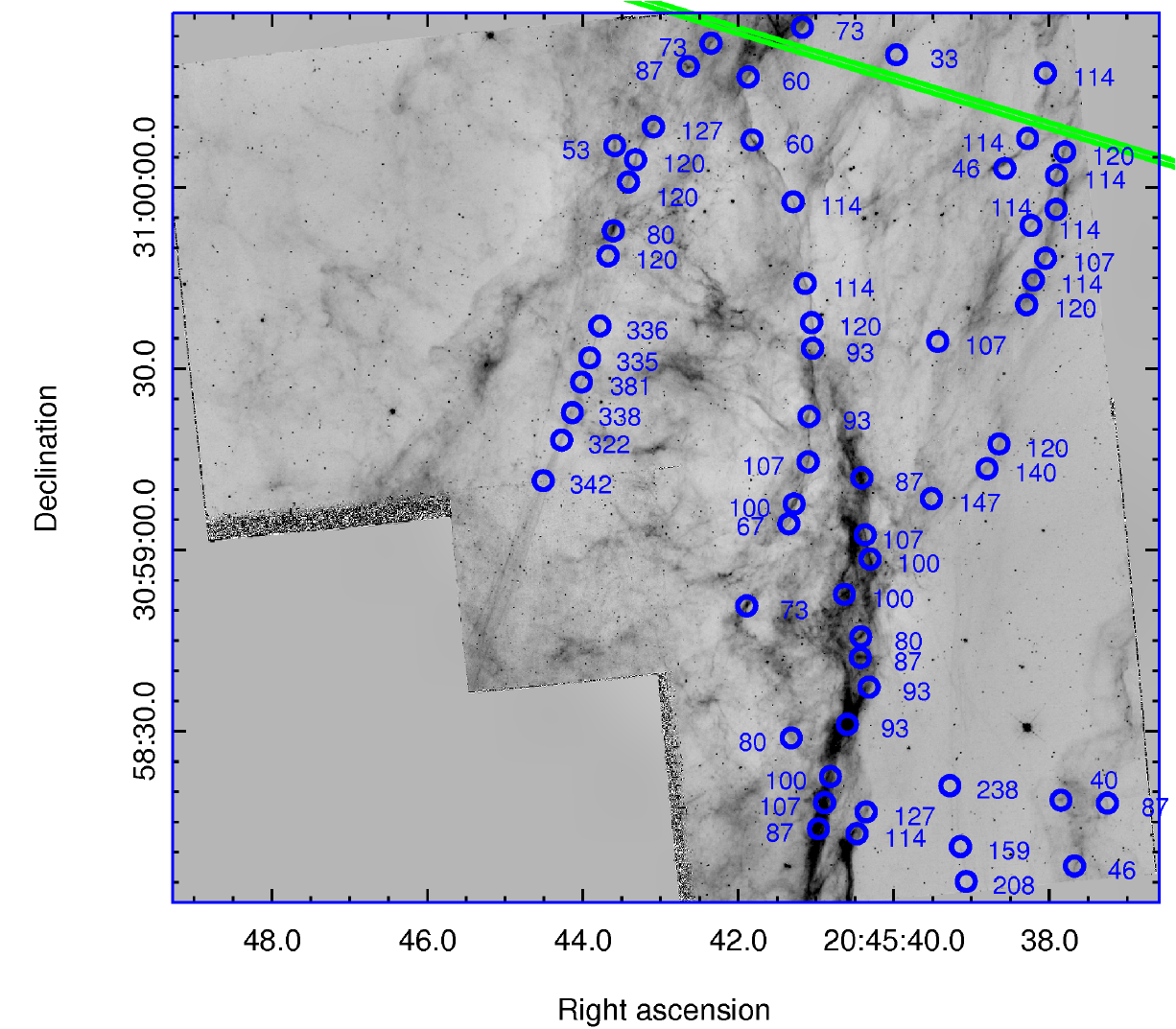}
\caption{Image in [O~III] (left) and H$\alpha$ (right) from WFPC2 in 1997.  This figure from Paper I shows the shock speeds for many filaments determined from the proper motions and the distance of 735 pc reported by Fesen et al. (2018).  The narrow green rectangles indicate the position of the Binospec slit that was used to study the structure along the flow direction.  It is parallel to, but slightly north of, the STIS slit as shown in Figures~\ref{o1_o2} and \ref{o3_ha}.
\label{o3_pm}
}
\end{figure*}

Proper motions were measured in Paper I by comparing the archival WFPC2 [O~III] image obtained by J. Hester in 1997 (Observing Program 5779) with the [O~III] image obtained as part of the Hubble Heritage program in 2015  (Observing Program 14056), and by comparing the 1997 H$\alpha$ image with the one we obtained on 2018 July 6 for this program.  We assumed the distance of 735$\pm$25 pc given by \citet{fesen18} to obtain shock speeds.  We reproduce Figure 3 from Paper I (our Figure ~\ref{o3_pm}) to show the range of shock speeds obtained.  

\begin{figure*}
\center
\includegraphics[angle=90,origin=c,trim={0cm 7cm 1cm 3cm}, clip,width=0.95\hsize]{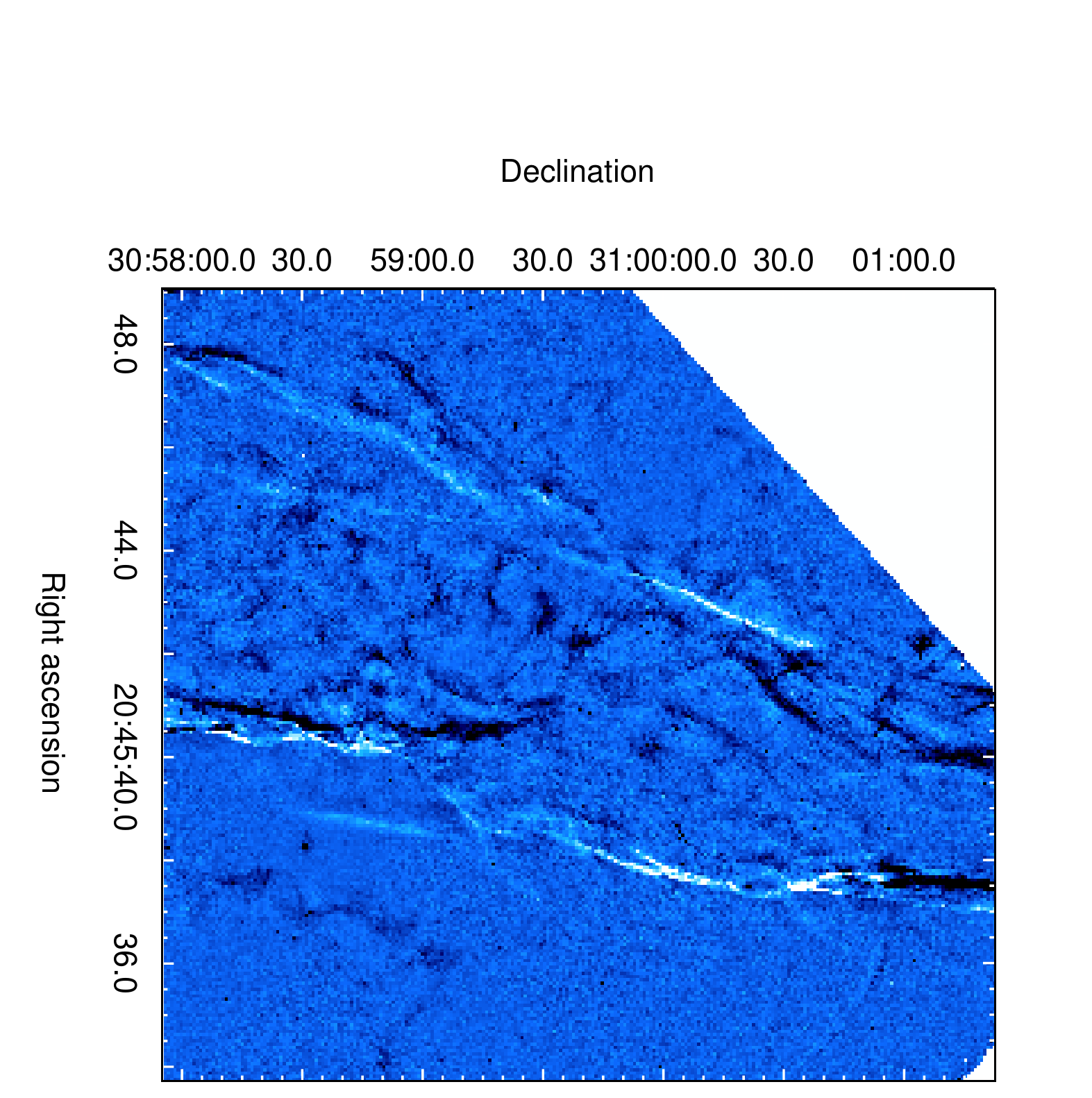}
\caption{Difference image of [O III] and H$\alpha$+[N II] from the Hubble Heritage project.  The [O III] image was multiplied by 1.2 to balance the images.  Bright [O III] emission is shown as white in this figure, while bright H$\alpha$+[N II] regions appear dark.
\label{diff_o3ha}
}
\end{figure*}

To highlight the contrasting morphologies of the [O III] emission and H$\alpha$, Figure~\ref{diff_o3ha} shows a difference image between the [O III] and the H$\alpha$+[N II] images from the Hubble Heritage project.  We use this pair to avoid features due to proper motion or changes in the emission than might occur for images taken at different epochs.  The image emphasizes the continuous nature of the [O III] filaments, which lie close to perpendicular to the shock motion, and the more fragmented appearance of the H$\alpha$ filaments, which often lie at large angles to the overall structure.

Paper I concentrated on the region observed with Binospec, as indicated by the slit shown in Figure~\ref{o3_pm}, which is located parallel to and slightly to the north of the STIS slit discussed below.  The full field shows a wider distribution of speeds.  The H$\alpha$ image shows a long, almost straight filament with speeds faster than 300 \kms\/ left of the center of the image.  This is a nonradiative shock similar to one observed somewhat to the north by \citet{raymond80}.  In such Balmer line shocks, H$\alpha$ photons are emitted by neutral H atoms that are swept through the shock and are excited before they are ionized.  This shock is projected on the emission from the slower shocks we are studying here, but it has little effect on any of the quantities we measure.  Analysis of this emission would require high-resolution spectra like those that \citet{medina14} obtained in the northern Cygnus Loop.

For the purposes of this paper, the most important feature of Figure~\ref{o3_pm} is the difference in shock speeds between [O~III] and H$\alpha$.  Many of the features seen in Figure~\ref{o3_pm} are apparent in [O III] or in H$\alpha$, but not both.  However, several long, clear filaments appear in both lines, such as the one seen in [O~III] at the top near the center, the one along the top right, and sections near the bottom center of the images.  In the north-central filament, the shock is 10 to 20 \kms\/ faster in [O~III] than in H$\alpha$, and in a few places this difference is up to 50 \kms.  In the northeast, the velocity difference ranges from zero up to 20 \kms .  In the southern filament H$\alpha$ is actually faster than [O~III] by 10 to 15 \kms .

\subsection{HST Spectra}

The set of STIS spectra obtained with the 52\arcsec\ slit centered at 20$^h$ 45$^m$ 37.853$^s, +30^{\circ}$ 59$^{\prime}$ 52\farcs49 (2000) is listed in Table 2.  We chose different slit widths for the UV and optical spectra.  Because of the higher sensitivity needed in the UV, we chose the 2\arcsec\/ slit, which provides a larger throughput, but has the disadvantage that the resulting lines are extremely broad, and even lines as far apart as He~II $\lambda$1640 and O~III] $\lambda$ $\lambda$ 1661, 1666 overlap to some extent.  We obtained optical spectra with the 0\farcs2 slit, but also spectra with the 0\farcs1 slit with the G430M and G750M gratings in order to obtain high spatial and spectral resolution to determine the ram pressure of the shock. All the spectra were obtained at a Position Angle of $73^{\circ}$.

The optical spectra suffered from the increasing noise level of the STIS CCD detectors due to changes in the readout noise, hot pixels, and charge transfer efficiency caused by radiation damage.  Even after the cosmic-ray rejection procedure, we find that we cannot combine as many pixels along the slit as we had planned, and therefore that the S/N is lower than expected.

\begin{table}
\begin{center}
\caption {STIS Spectra}
\begin{tabular}{l l l c }
\hline
\hline
Spectrum   &  Grating  &  Slit  & Exposure \\
\hline
0503010  &  G140L  &  2\arcsec  &  2220 \\
0503020  &  G240L  &  2\arcsec  &  2583 \\
0503030  &  G430L  &  0\farcs2 &   2649 \\
0504010  &  G750L  &  0\farcs2 &   2329\\
0504030  &  G430M  &  0\farcs1 &  2684 \\
0504040  &  G750M  &  0\farcs2 &   2612 \\
\hline
\end{tabular}
\end{center}
\end{table}

\begin{figure}
  \center
\includegraphics[trim={2 0 0 0},clip,width=0.50\hsize]{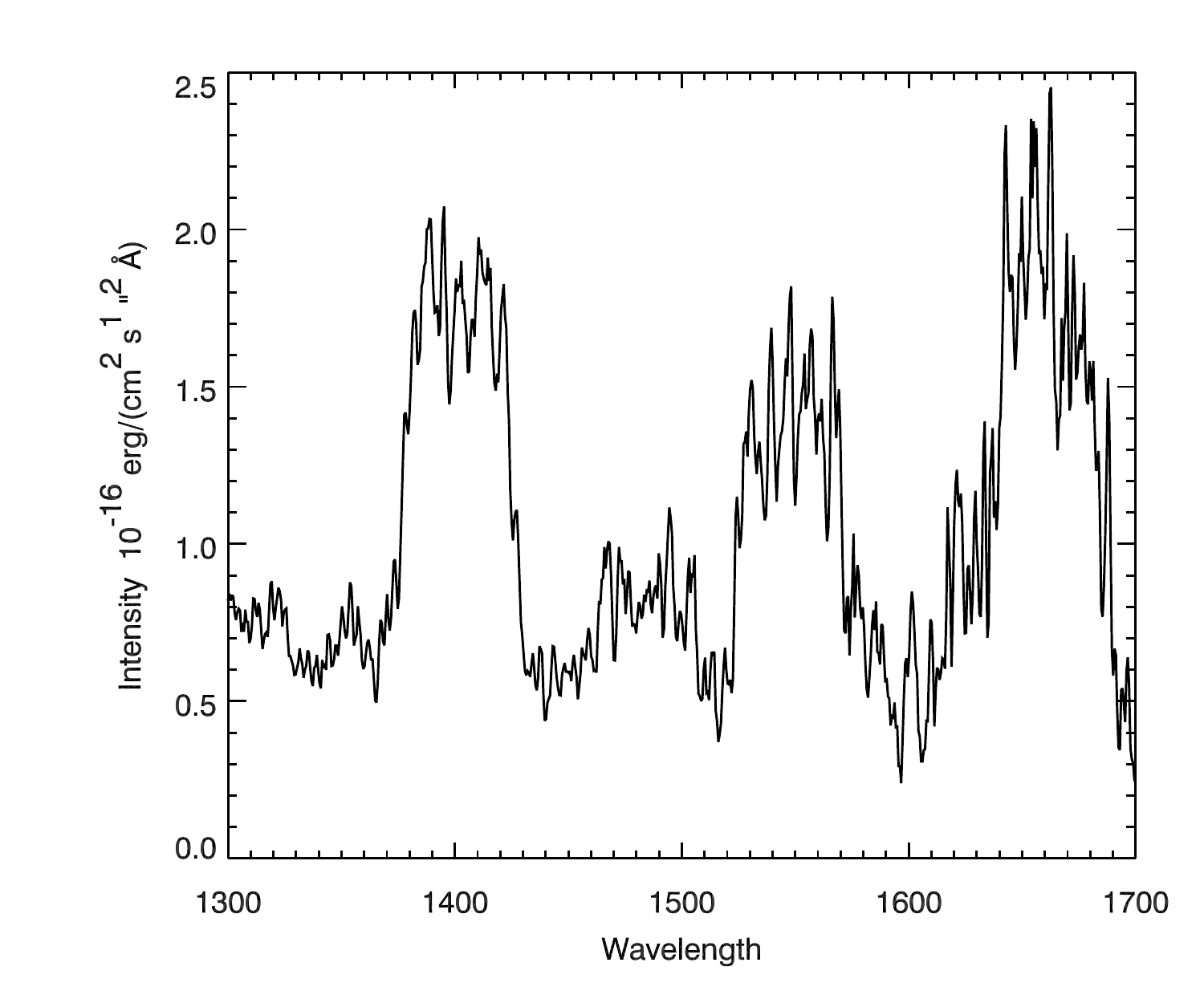}
\includegraphics[trim={2 0 0 0},clip,width=0.50\hsize]{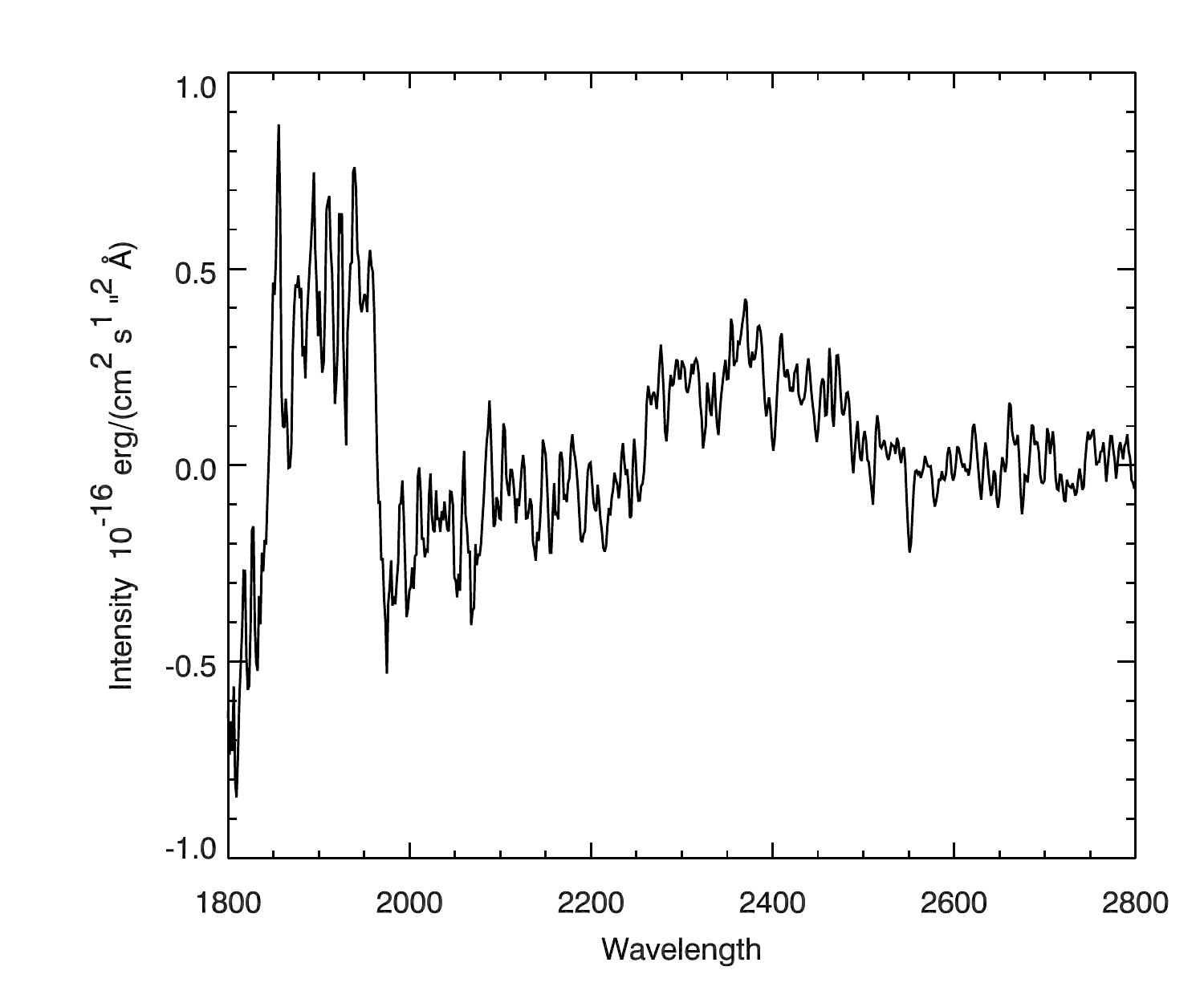}
\caption{STIS UV spectra averaged over the 3\farcs7 spatial band where the emission is brightest.  The FUV spectrum is shown on the top, and the NUV on the bottom.  The 2\farcs0 slit gives a high count rate, but it makes the lines extremely wide. 
\label{uvspec}
}
\end{figure}  
  
Figure~\ref{uvspec} shows plots of the G140L and G240L spectra averaged over a 3\farcs7 band extending from -2\farcs288 to 1\farcs402 from the reference position.  This is the brightest region in [O III], and it accounts for nearly all the flux in the UV lines.  The FUV lines are brighter toward the eastern end of the extracted region, while the NUV lines are brighter toward the west.  Note the blend of He II and O III] near 1650 \AA \/ and the blend of C II]+Si II] with [Ne IV] near 2390 \AA .  In both cases, there is a peak where the lines overlap with relatively flat shoulders on both sides.  We modeled the  blends as the sums of top-hat profiles centered at the wavelengths of the strongest contributors, thus allowing individual line fluxes to be derived.  



\begin{table}
\begin{center}
\caption {Line Intensities}
\begin{tabular}{l l r r r}
\hline
\hline
$\lambda$   &  Ion  &  F\tablenotemark{a} & I\tablenotemark{b} & Model \\
\hline
1400  &  O IV], Si IV  & 8.7   &  3140 & 3050\tablenotemark{c}\\
1486  &  N IV]   & 2.00   &   661 & 540 \\
1550  &  C IV    & 5.52  &   2090 & 9100\tablenotemark{c} \\
1640  &  He II   & 3.49  &   1052 & 428 \\
1664  &  O III]  & 5.98  &   1800 & 1720 \\
1909  &  C III]  & 6.10 &   1950 & 2380 \\
2325  &  C II],Si II]  & 2.85 &   203 & 1100 \\
2421  &  [Ne IV] & 1.97  &  570  & 637 \\
3727  &  [O II]  & 9.08  &   1560 & 3760 \\
5007  & [O III]  & 12.5  &   1684 & 2530 \\
6300  & [O I]    & 0.50  & 57  & 16 \\
6563  & H$\alpha$& 2.60  & 300 & 300 \\
6584  & [N II]   & 1.50   & 170 & 467 \\
6725  & [S II]   & (3.25) & (363) & 516 \\

\hline
\end{tabular}
\end{center}
\tablenotetext{a}{$10^{-15} \rm erg~cm^{-2}~s^{-1}~arcsec^{-2}$.} 
\tablenotetext{b}{Corrected for E(B-V) = 0.2, scaled to H$\alpha$=300.}
\tablenotetext{c}{The C~IV line and the Si~IV part of the 1400 \AA\/ blend (10\%) are attenuated by resonant scattering.}

\end{table}

Table 3 gives the resulting line intensities and the relative intensities corrected for a reddening E(B-V) = 0.2 determined from the Binospec spectra in Paper I with the \citet{cardelli89} extinction curve.  This reddening is higher than is typical for the Cygnus Loop, but it is consistent with the dark cloud on the western edge of the remnant \citep{fesen18}.  Our optical STIS spectra used narrower slits and therefore do not pertain to exactly the same spatial region. Therefore, we use averaged intensities from the WFC3 images from the region corresponding to the STIS 2\farcs0 slit for the optical lines.  For [N~II] $\lambda$ 6584 we used the difference between the H$\alpha$+[N~II] image from the Hubble Heritage project and our narrow H$\alpha$ image. The model shown in Table 3 is discussed below.

\section{Analysis}

\subsection{Images}

We apply a number of techniques to the images in order to quantify the flow properties of the shocked gas.  The ratio of the [O~II] image to the H$\alpha$ image yields the average temperature where those lines are formed.  We examine cross correlations and power spectra to investigate the scale sizes.  We also use the Rolling Hough Transform \citep{duda72,clark14} technique to bring out linear features in the images, allowing us to quantify the shape of the ripples, compare H$\alpha$ and [O~III], and measure variations in proper motions along individual filaments.  
  
\subsubsection{Temperature map} 

\begin{figure*}
\center
\plotone{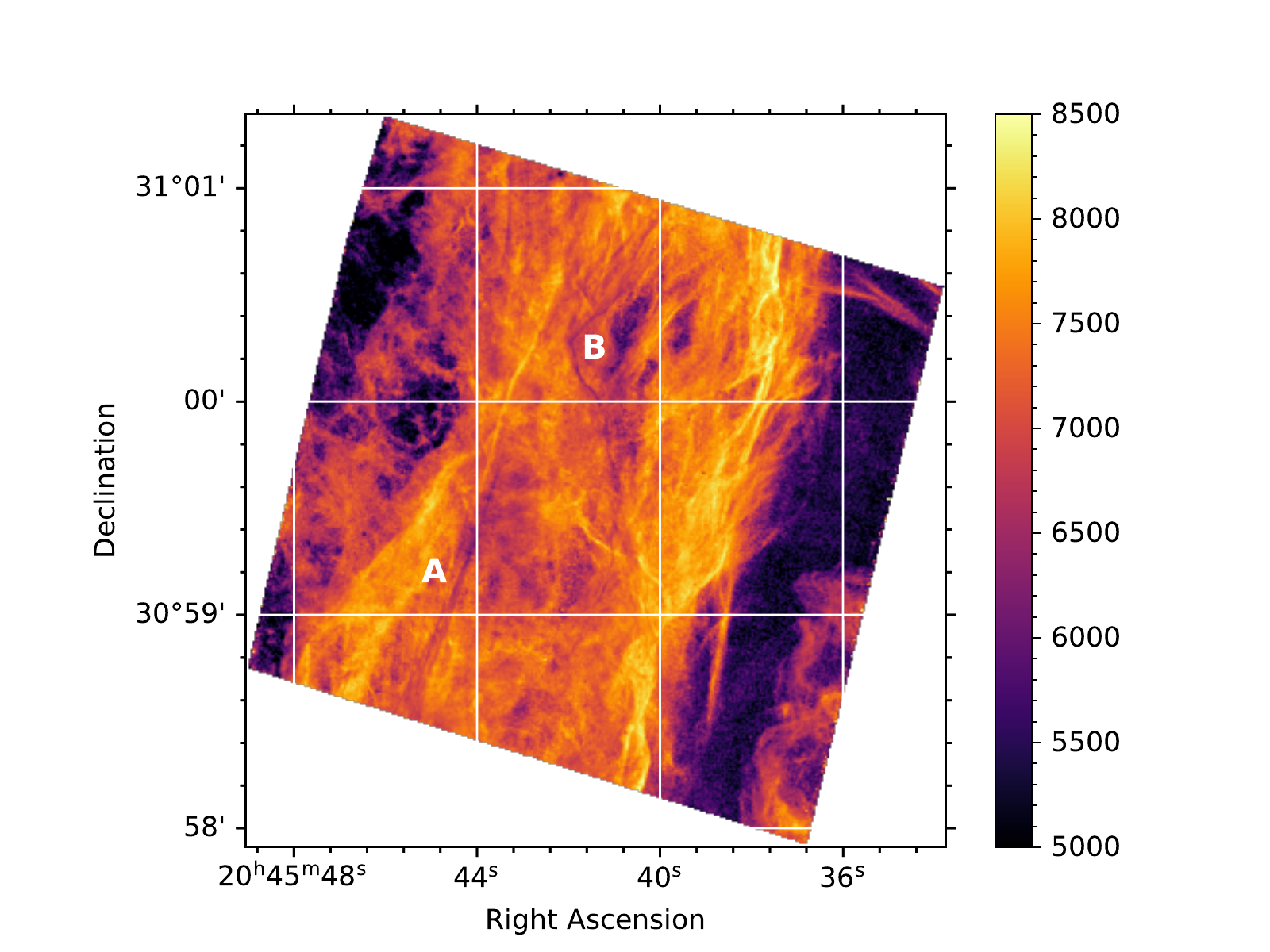}
\caption{Temperature obtained from the ratio of the [O II] and H$\alpha$ images.  Most dark regions have too few counts for a reliable temperature estimate, but the straight, narrow filament at 20$^h$ 45$^m$ 44$^s$ and position angle -15$^\circ$ marked  A is the nonradiative shock that emits only Balmer lines.  The curved filament marked B that angles toward the NE, then turns back to the NW is also a real feature. The images were binned to 4x4 pixels (0\farcs16) to improve the signal-to-noise ratio.}
\label{T_o2ha}
\end{figure*}

The ratio of [O II] $\lambda$3727 to H$\alpha$ provides a temperature estimate.  Recombination rates scale roughly as T$^{-1/2}$ while collisional excitation rates scale roughly as T$^{-1/2}$ exp(-E/kT).  Therefore, the ratio of a collisionally excited line to a recombination line can be used to estimate the temperature.  In Paper I we used the [N~II] to H$\alpha$ ratio from the Binospec spectra to obtain temperatures of 7000 to 9000 K in the region where those lines are formed.  Here we use the [O~II] to H$\alpha$ ratio from the WFC3 images.  The oxygen ionization state is tightly tied to that of hydrogen by rapid charge transfer.  \citet{fieldsteigman} showed that n(O$^+$/n(O$^0$) = (8/9) n(H$^+$)/n(H$^0$).  Therefore, the formation regions of the two lines are virtually the same, except for a small amount of H$\alpha$ from regions where oxygen is ionized to O~III or above. The [O~II] line has the advantage over [N~II] of a higher excitation potential, making it more sensitive to temperature, and the disadvantage that a significant reddening correction is required.  As in Table 3, we use E(B-V)=0.2 with the \citet{cardelli89} reddening to obtain a correction factor of 1.51 for the [O~II] to H$\alpha$ ratio.

We use [O~II] excitation rates from CHIANTI Version 8 \citep{dere97, delzanna15} and H$\alpha$ Case B recombination rates from \citet{hummerstorey}.  An oxygen abundance must be assumed, and as in the shock models computed here and in Paper I, we take 8.82. Under the assumption that the H$\alpha$ is produced entirely by recombination, we find

\begin{equation}
    [O~\mathrm{II}]/H\alpha = 130 ~(T/10000)^{0.48} ~e^{-38560/T}
\end{equation}

\noindent
in photon units.  The power-law term arises from departures of both the [O~II] and H$\alpha$  emissivities from the simple T$^{-1/2}$ dependence mentioned above.  We can approximately invert the relation to obtain

\begin{equation}
T/10000 \simeq 4.8/(0.9 - ln(O~\mathrm{II}/H\alpha) / 146).
\end{equation}

An assumption of this method is that collisional excitation is a minor contribution to the H$\alpha$ flux.  In collisional ionization equilibrium, the collisional excitation of H$\alpha$ can dominate, but in radiative shock waves both time-dependent ionization and photoionization reduce the neutral fraction.  We do not have a model-independent way to determine the collisional contribution to H$\alpha$, but we can estimate the collisional component from the models presented in Paper I and in Table 3.  They show that the collisional contribution is small, and hence the formula above is reliable up to a temperature of about 12,000 K, above which the formula underestimates the temperature.  

Figure~\ref{T_o2ha} shows the temperatures from the ratio of the WFC3 [O~II] and H$\alpha$ images.  Regions where the signal was too low to provide reliable temperatures are shown as very dark.  The temperatures range from below 5000 K to just above 9,000 K.  They  are similar to those obtained in Paper I from the [N~II] to H$\alpha$ ratio, but they show a wider range because the images sample a larger region at higher resolution. They were binned to 4x4 pixels (0\farcs16) to increase the signal to noise.  Recall that these temperatures are averages of the regions where those two lines are formed, and that they could be underestimated above 12,000 K.  

\begin{figure*}
\center
\includegraphics[trim={2cm 0 0 0},clip,width=0.45\hsize]{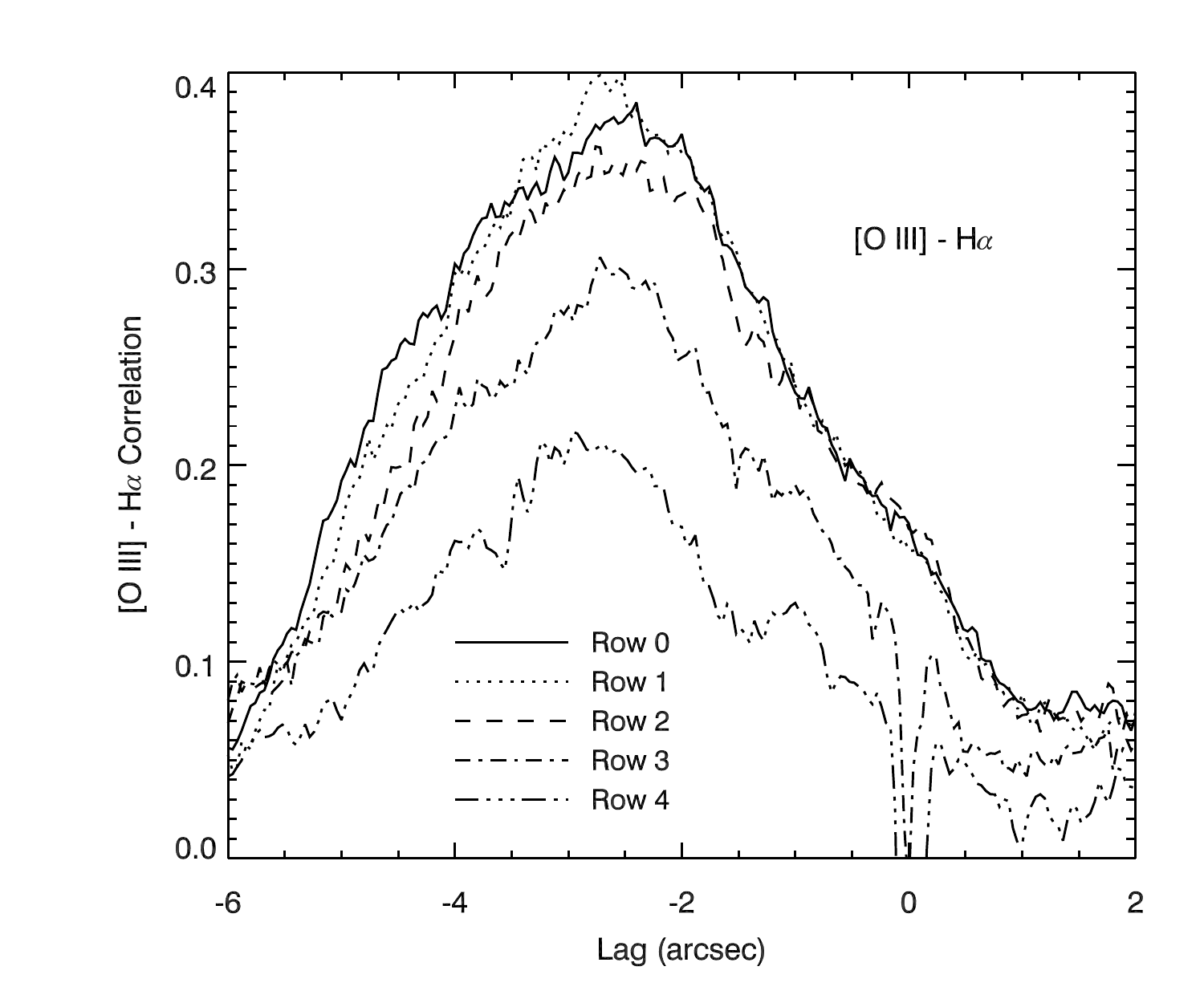}
\includegraphics[trim={0 0 0.5cm 0},clip,width=0.5\hsize]{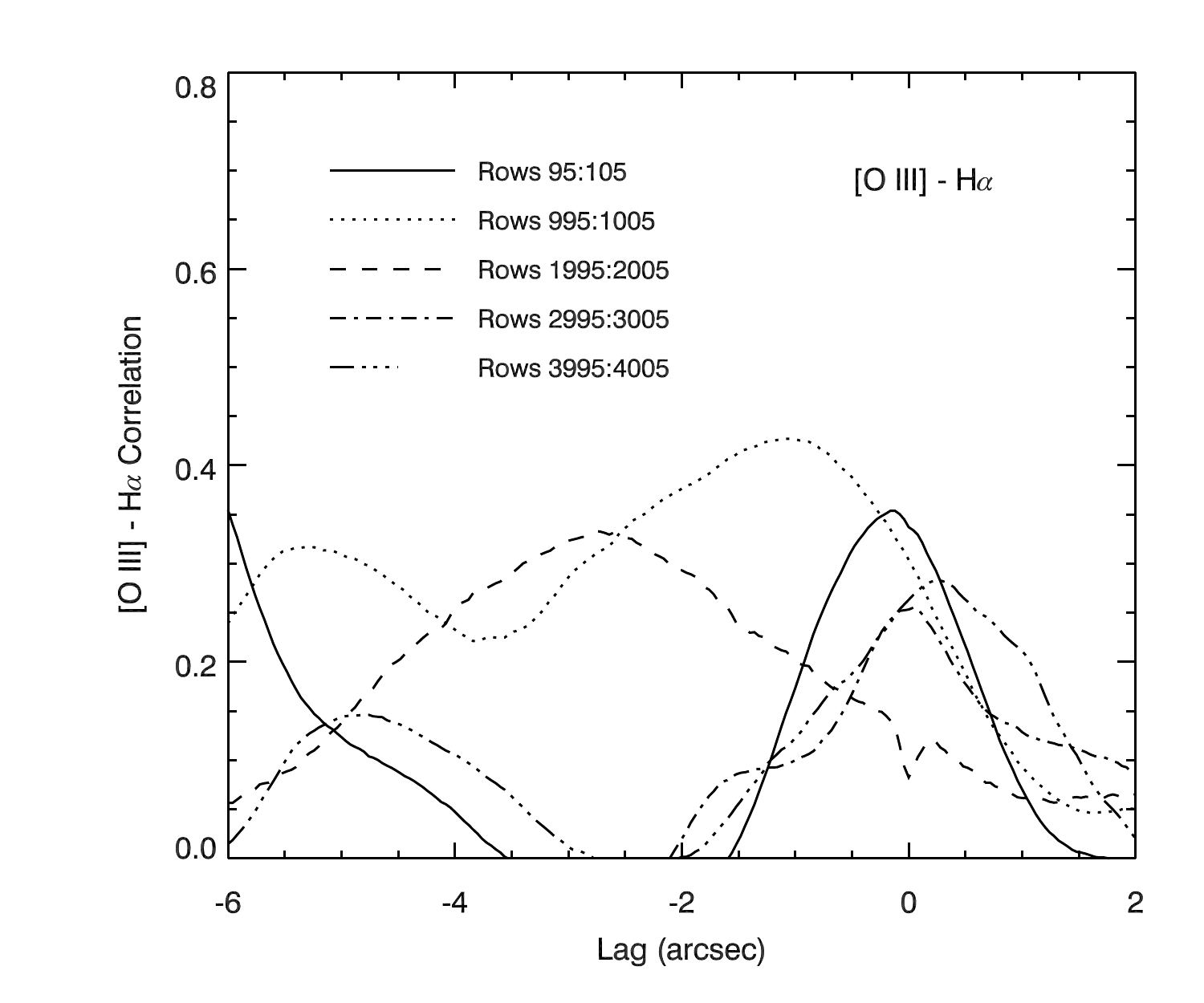}
\caption{Cross correlations along the flow direction. Left panel;  Cross correlations between [O~III] and H$\alpha$ in 5 adjacent rows along the shock direction.  There are differences among these rows, though they are separated by only 0\farcs04, but the shapes are the same except for the subtracted star in some rows that produces a dip at 0. Right panel; Average cross correlations for sets of eleven rows extracted at positions from near the bottom to near the top of the region being analyzed. 
\label{xcorr_o3_ha}
}
\end{figure*}

Comparison of the temperature map with the separate [O~II] and H$\alpha$ images indicates that the bright [O~II] filaments along the eastern and western sides lie in the 8000-9,000 K range, while lower temperatures are found in the H$\alpha$-bright regions in the central regions, as well as in regions trailing some of the [O~II] filaments.  The bright filaments, especially those along the western edge, presumably show higher temperatures because the H$\alpha$ emission is formed in a thicker recombination zone, while the [O~II] emission is strongly weighted to the higher temperatures where the plasma is just beginning to recombine.

A few low temperature filaments are visible in the image.  The very narrow, straight filament labeled 'A' in Figure~\ref{T_o2ha} is a nonradiative shock (Balmer line filament) discussed above.
There is also a more irregular filament in the north-central part of the image marked 'B' in Figure~\ref{T_o2ha}.  Temperatures there fall below 6,000 K, suggesting the formation of a cold shell.  However, two other interpretations are possible.  We have assumed a uniform reddening for the entire region observed, but the H$\beta$ to H$\alpha$ ratio shows some variation (Paper I). It is possible that a thin sheet of swept-up dust is seen edge-on, and that it attenuates the [O~II] emission along the line of sight.  Shocks are believed to shatter dust particles by grain-grain collisions as the gas is compressed \citep[e.g.]{jones96, slavin15}, and smaller particles might increase the extinction at short wavelengths.  Another possibility is that this filament is actually a relatively slow shock, as indicated by speeds around 60 \kms\/ in Figure~\ref{o3_pm}.  A slow shock in partially neutral gas can produce very strong H$\alpha$ emission by collisional excitation \citep{coxraymond85}, in which case the temperature is grossly underestimated.  We do not have sufficient information disentangle these different scenarios.

\subsubsection{Autocorrelations and cross-correlations}

In order to perform cross-correlation analyses parallel and perpendicular to the general flow direction, we rotated the images by 17$^{\circ}$, matching the direction of the Binospec slit shown in Figure~\ref{o3_pm} and the STIS slit shown in Figures~\ref{o1_o2} and ~\ref{o3_ha}.  Figure~\ref{xcorr_o3_ha} shows the cross-correlations for 5 adjacent rows (0\farcs04 offsets) along the flow direction in the left panel.  They show some difference in amplitude, but the same general  shape.  The spike at zero is caused by a star in some of the rows.  Overall, the left panel is what would be expected from the 1D model described in Paper I if the structure on the plane of the sky is mainly due to projection of a gently rippled sheet seen nearly edge-on \citep{hester87}.  The [O~III] emission leads the H$\alpha$ by 2\arcsec\ to 3\arcsec, and the width of the correlation is around 4\arcsec .  

On the other hand, the right panel of Figure~\ref{xcorr_o3_ha} shows that the agreement with the picture of a 1D shock perturbed by gentle ripples is not always the case.  Here we see averages of groups of 11 rows distributed from near the bottom to near the top of the observed region.  While two of the curves show broad peaks displaced by 1\arcsec\ to 3\arcsec\ from zero, the other three do not.  Two show strong peaks near zero offset, suggesting either coincidental superposition of shocks mainly seen either in H$\alpha$ or in [O~III], or else structure unrelated to the cooling flow, such as strongly compressed regions that emit in both lines.
  
\begin{figure*}
\center
\includegraphics[width=1.00\hsize]{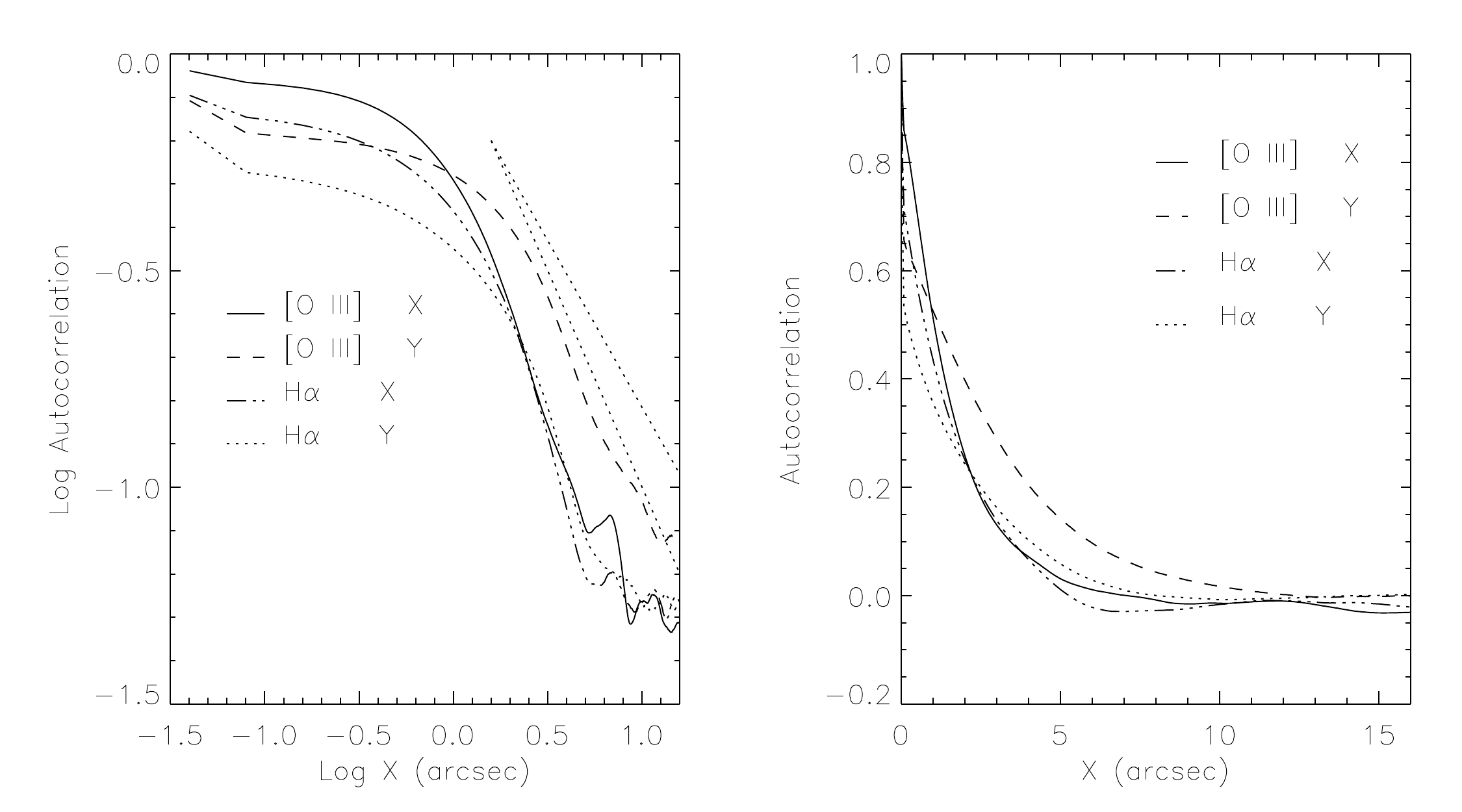}
\caption{Autocorrelation functions of [O~III] and H$\alpha$ intensities in the east-west (X) and north-south (Y) directions.  The left panel presents the data on a log-log scale.  For that plot the average of the absolute values of the autocorrelations are shown.  The right panel presents the average autocorrelations (not the absolute values) on a linear scale.  The straight dotted line in the left panel indicates a slope of -1.}
\label{acorr}
\end{figure*}

The autocorrelation functions in different emission lines indicate the scale sizes of the emitting structures.  Figure~\ref{acorr} shows the autocorrelation functions of the [O~III] and H$\alpha$ lines averaged over a 4096x4096 box in the south-central section of Figure~\ref{o3_ha}, chosen to avoid the regions where there is no data and the regions to the west that have not yet been reached by the shock. We used the Heritage [O~III] and H$\alpha$+[N~II] images for consistency, and stars were mostly removed before the autocorrelations were computed. The X- (east-west) and Y- (north-south) directions are close to parallel and perpendicular to the shock flow.  The log-log presentation in the left panel is the average of the absolute values of the autocorrelations of the 4096 rows or columns, while the linear plot on the right is the direct average of the 4096 rows or columns.  

Figure~\ref{acorr} shows a very steep drop in the autocorrelation on scales of 1\arcsec\ to 2\arcsec, which is the typical thickness of the filaments in both [O~III] and H$\alpha$.  The secondary peaks on the [O~III] autocorrelation in the X direction indicate that filaments often occur in pairs due to the rippled structure of the shock.  The autocorrelation of the [O~III] emission in the Y-direction is stronger at scales extending to about 5\arcsec\ because the filaments are extended in roughly the north-south direction.  

\subsection{Fourier Power Spectra}

\begin{figure*}
\begin{centering}
\includegraphics[trim={2cm 0 0 0},clip,width=0.45\hsize]{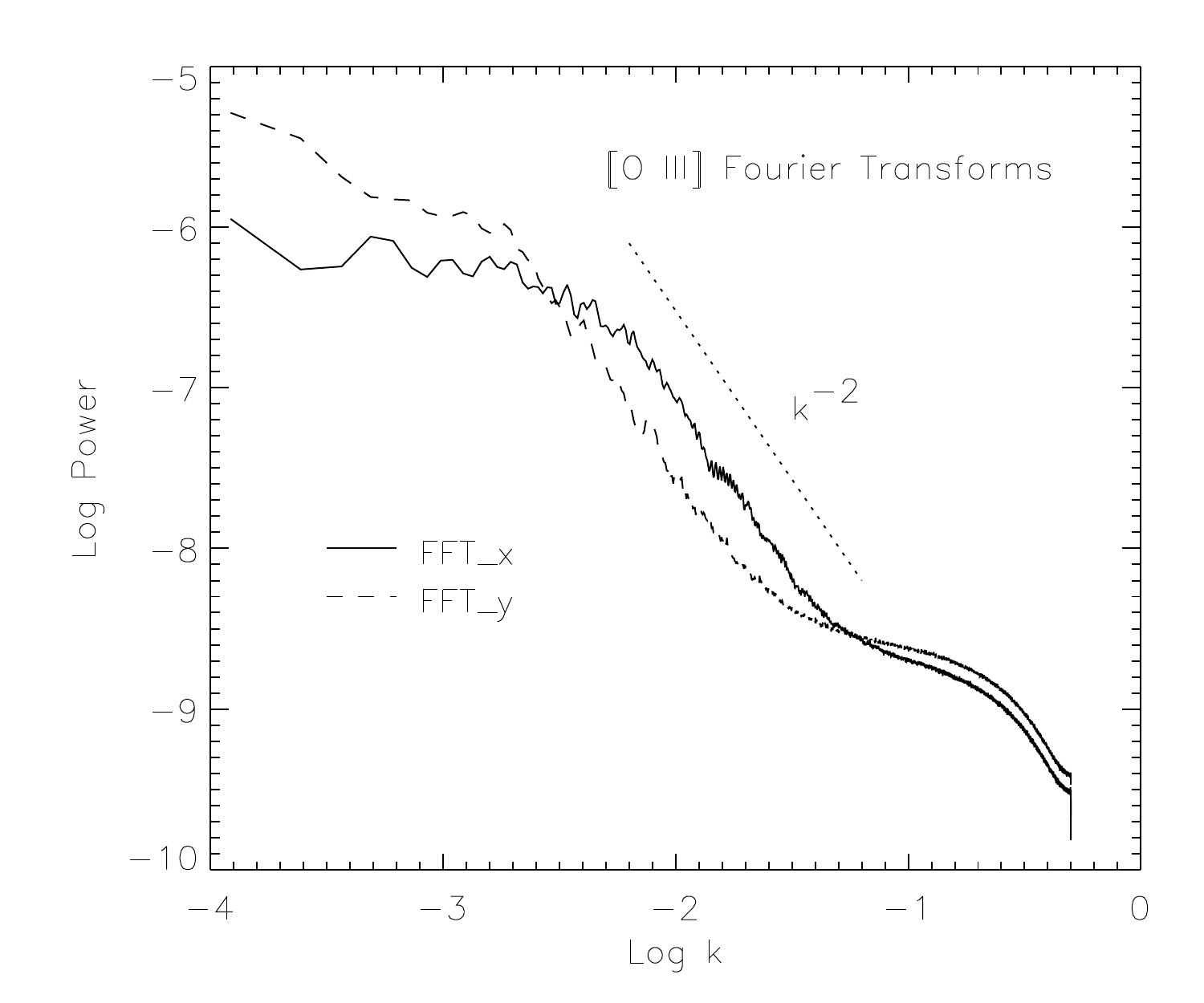}
\includegraphics[width=0.51\hsize]{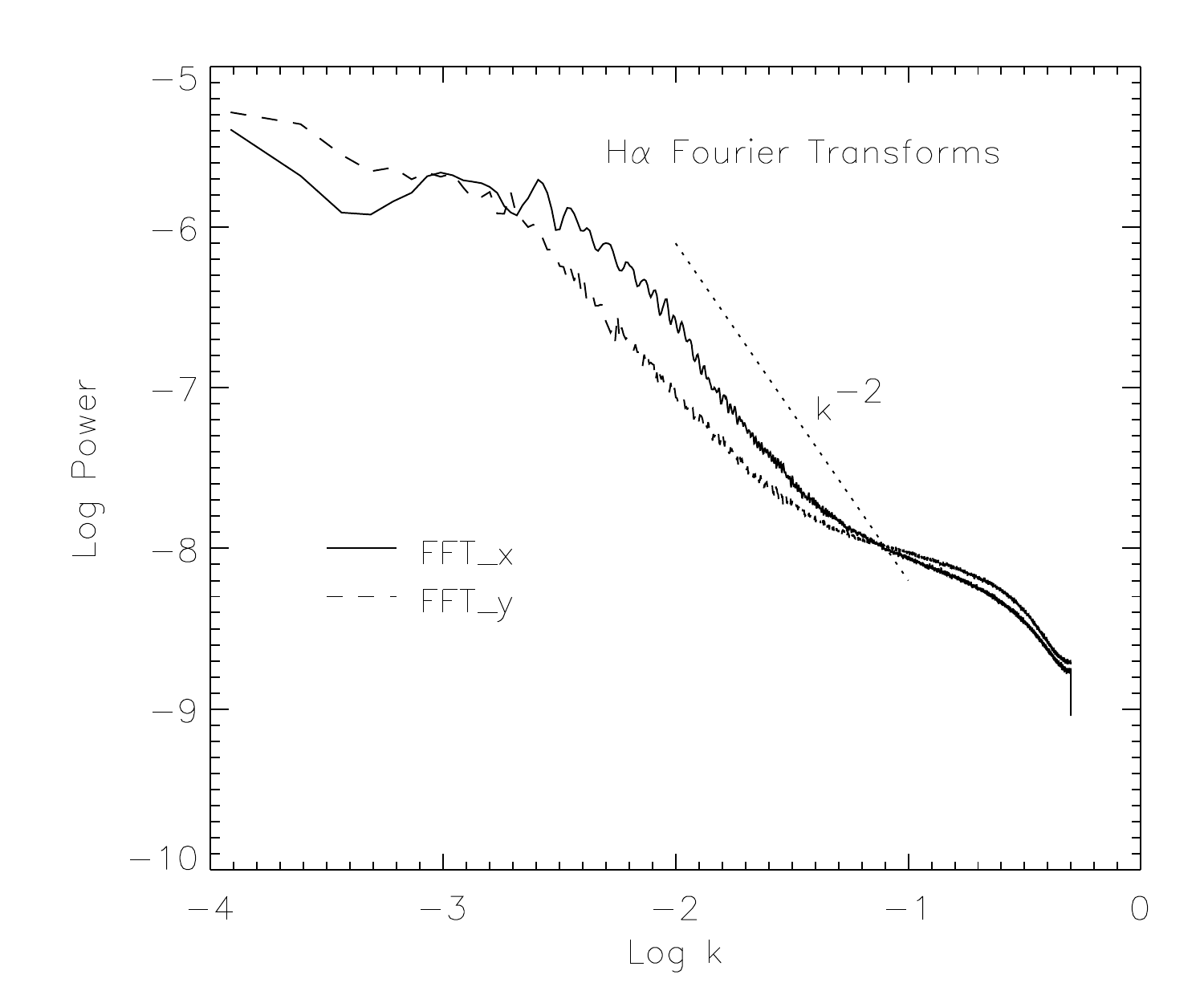}
  \caption{Fourier power spectra of the central regions of the [O~III] and H$\alpha$+[N~II] images in the X- (east-west) and Y- (north-south) directions as functions of the wavenumber, k.  The flat sections at higher frequencies may be dominated by faint stars that were not removed from the images.}
  \label{ffts}
  \end{centering}
  \end{figure*}

Power spectra obtained from Fourier transforms are commonly used to characterize turbulence in space plasmas.  Figure~\ref{ffts} shows the average Fourier power spectra obtained from the central sections of the [O~III] and the H$\alpha$+[N~II] images from the Heritage data.  Stars were mostly removed from the images, but some remain because removing the fainter stars would have removed some of the signal.  The flat regions at fairly high wavenumber, $k$, in both plots result from the general noise level or from unremoved stars, and the drop-off at the highest k may indicate the size of the stellar images. 

The Fourier transforms in the X- and Y-directions are close to parallel and perpendicular to the shock direction, so the X- and Y- Fourier transforms show structure along and perpendicular to the overall shock flow.  The differences in power between the X- and Y-directions reflect the elongation of filamentary structure mainly close to north-south.  The spectra are generally close to $k^{-2}$ power laws in the range that might be interpreted as a turbulent cascade.  The H$\alpha$ power spectrum in the Y-direction is shallower than that in the X-direction, suggesting less intermediate-scale structure.

As discussed in Paper I, the filaments are tangencies of a rippled sheet of emitting gas to the line-of-sight (LOS) (Hester 1987).  This is especially clear in the [O~III] image, and it is supported by the coincidence of bright filaments with zero Doppler velocity (Paper I).  In that picture, the brightness corresponds to the depth of the tangency of the LOS to the ripple. Therefore, a Fourier transform of the brightness provides information about the ripples, but it is very different from the Fourier transforms of velocity, density or magnetic field fluctuations that are generally measured in turbulence studies.

A toy model of the rippled sheet is a sine wave of given amplitude and thickness.  The number of pixels within the sheet as a function of distance from the axis simulates the brightness when viewed end-on.  The Fourier transform of that brightness shows a $k^{-1}$ slope at very low wavenumbers and a break to $k^{-2}$ at intermediate wavenumber, very much like the observed power spectra in Figure~\ref{ffts}.  At high wavenumbers, the FFT of the toy model varies so violently that is is hard to assign a slope.  The wavenumber of the break from a slope of -1 to -2  depends on the amplitude of the ripple and the thickness of the sheet, increasing with decreasing thickness and decreasing amplitude.

The basic similarity of the H$\alpha$ and [O~III] power spectra suggests that some of the same basic ripple structure is seen in the two lines, though the visual impression of the H$\alpha$ image is dominated by the more diffuse emission.  This is confirmed by an assessment of the Rolling Hough Transforms discussed below.  It also suggests that the overall brightness is governed mostly by geometry (tangencies of the LOS to ripples) rather than density, though the shallower slope of the FFT of the H$\alpha$ image in the Y-direction suggests more power in small features or else a non-geometrical contribution to the H$\alpha$ brightness.

  
 \subsection{Rolling Hough Transform Analysis}
 
The Rolling Hough Transform (hereafter RHT) picks out linear features in an image.  It was introduced by \citet{duda72} and later used by \citet{clark14} for the analysis of magnetically aligned linear features in radio H~I data.  We applied the RHT {\sc python} code developed by Clark et al. (available at \url{github.com/seclark/rht}) to the HST images in order to quantify the amplitude of ripples in the shock and to compare the level of turbulence in the [O~III] and H$\alpha$ zones of the cooling region.  The code uses unsharp masking to bring out thin features, then calculates the sum of the brightness along lines through each point as a function of position angle.  The position angle of a line has a 180 degree degeneracy, and we have wrapped the angles around 180 degrees when necessary.  We binned the data to 5x5 pixels (0\farcs2) to reduce the noise, then we chose the parameters smooth = 15 and xleng = 41 for the smoothing length and filament sample length, respectively.
 
 \begin{figure*}
\centering
\plottwo{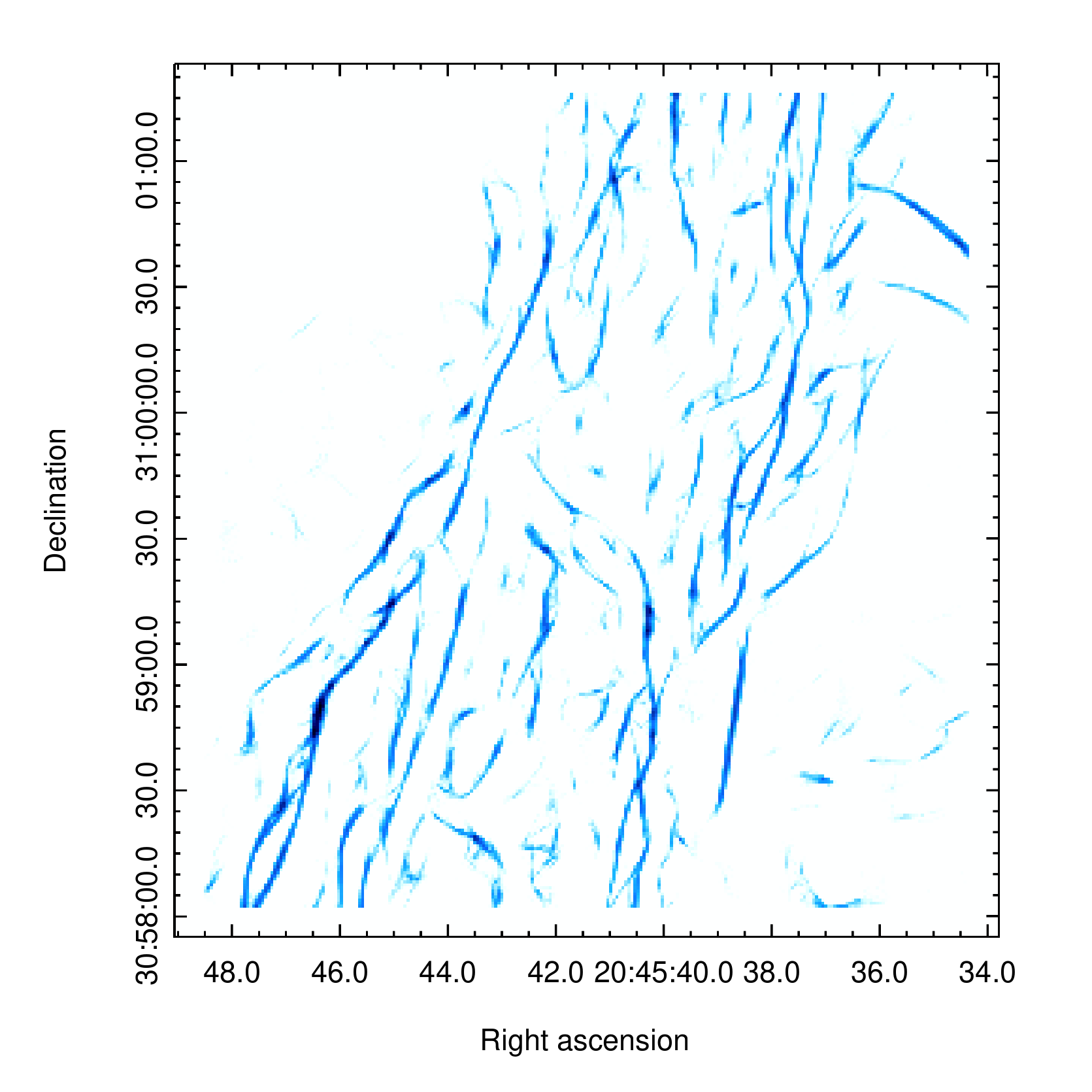}{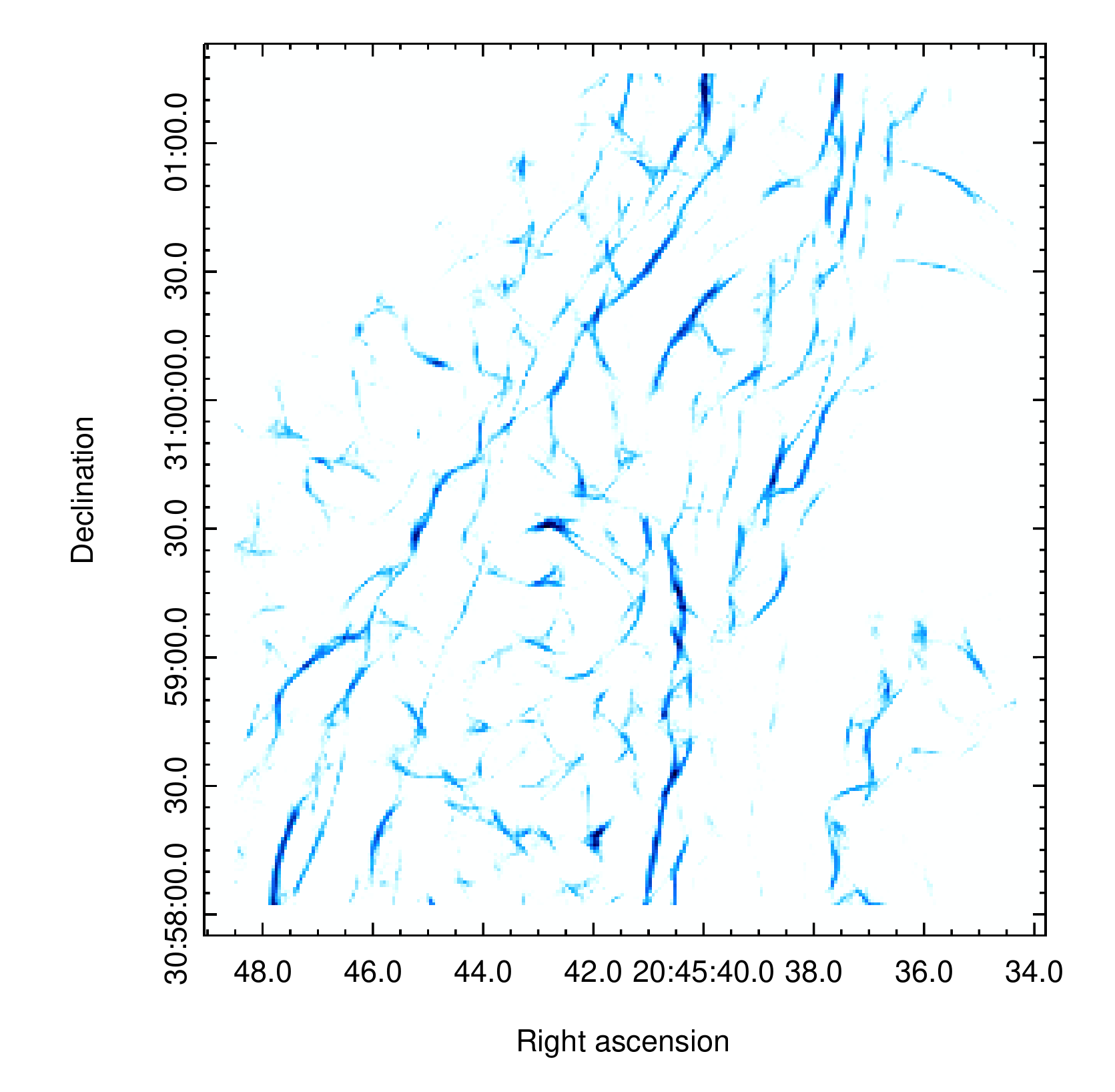}
\caption{Left and right panels; Rolling Hough Transforms of sections of the [O~III] and H$\alpha$ images.  The data used for these images were binned to 5x5 pixels (0\farcs2)}     \label{rht_o3ha}
 \end{figure*}
 
 \begin{figure*}
\centering
\includegraphics[trim=0cm 0cm 0cm 0cm,clip,width=0.8\textwidth]{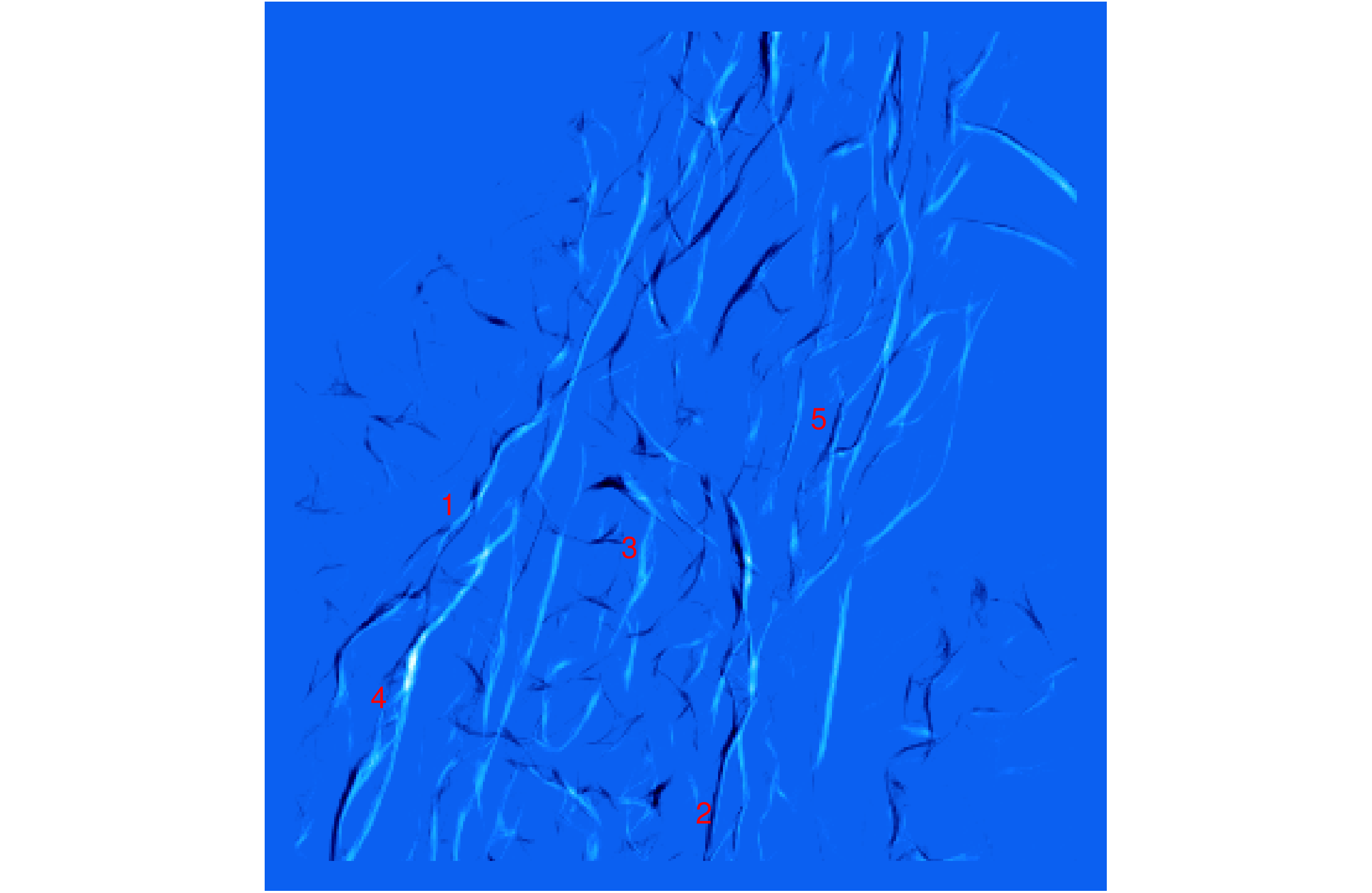}
\caption{Difference between RHTs of [O~III] (white) and H$\alpha$ (black).  The [O~III] filaments lead the H$\alpha$ filaments where both lines can be seen, but many features appear in only one line.  The image is 196\arcsec x 208\arcsec as in Figure~\ref{rht_o3ha}.  The labels indicate 1) a rare instance of an H$\alpha$ filament preceding the [O III] filament and 2) a long, nearly NS filament discussed below.  Labels 3), 4) and 5) indicate small, medium and large "eye-shaped" features discussed below.}
\label{rht_diffo3ha}
 \end{figure*}

 \subsubsection{[O~III] - H$\alpha$ separation from RHTs}
 
Figure~\ref{rht_o3ha} shows the RHTs of the [O~III] and H$\alpha$ images, and Figure~\ref{rht_diffo3ha} shows the difference between the two.  Several fairly long filaments show up in both lines with H$\alpha$ trailing [O~III], but other filaments show up only in one of the lines or fade from one line to the other along their length.  While our parameter choices for the RHT will affect which filaments show up, a clear difference is observed.  This is likely to result from shock speed and shock completeness.  Shocks slower than 100 \kms\ produce little [O~III] unless the oxygen in the preshock gas has been photoionized to O$^{2+}$ \citep{hrh87}, while faster shocks that have not had time to cool to 10,000 K will show up only in [O~III] \citep{raymond88}. 
 
 \begin{figure}
 \centering
 \includegraphics[width=0.7\textwidth]{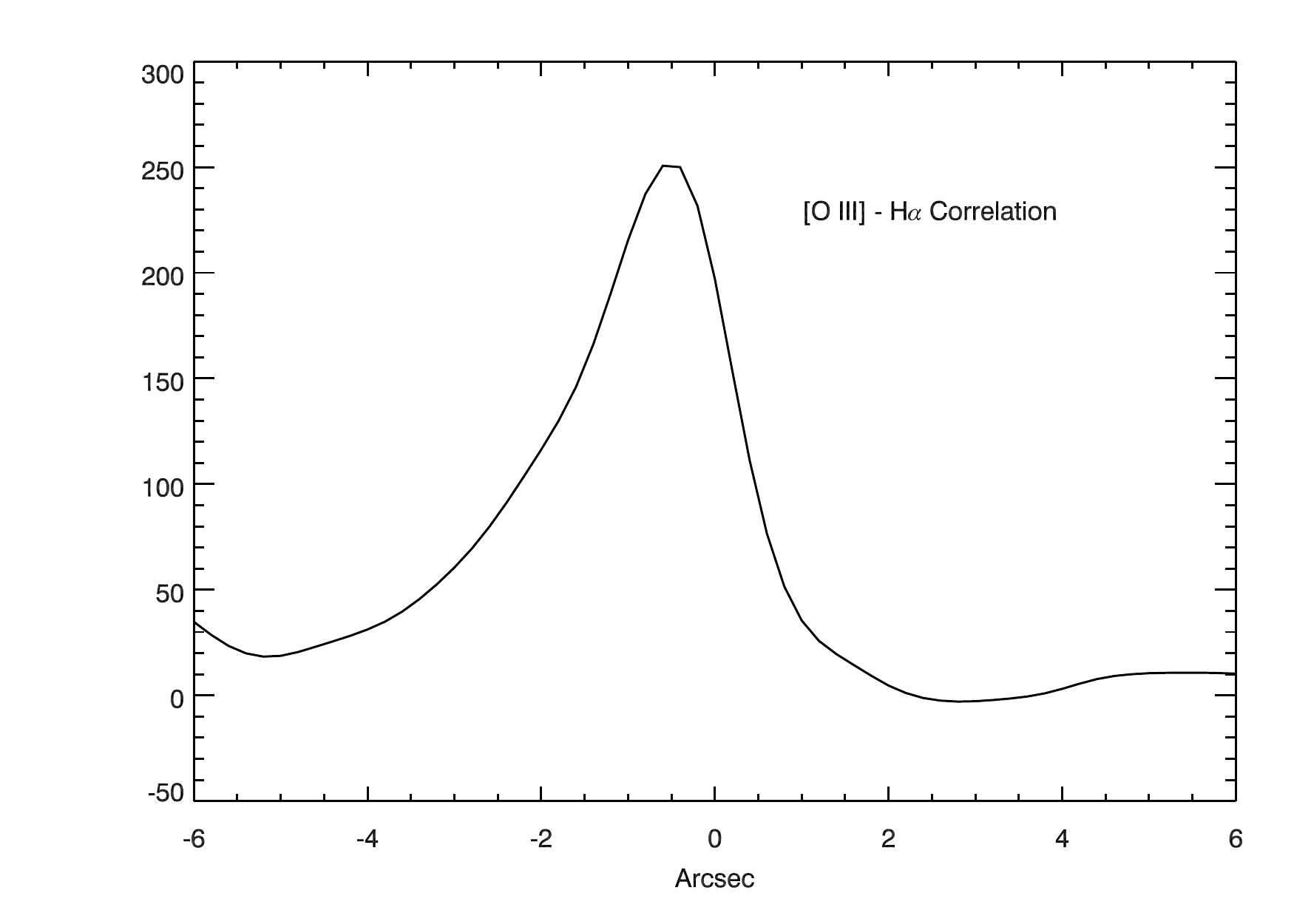}
 \caption{Cross correlation in the east-west direction between [O~III] and H$\alpha$ RHT images.  The H$\alpha$ peaks about 0\farcs5 behind the [O~III] and extends to about 2\farcs0.}
 \label{ccor_rht}
 \end{figure}
 
Cross-correlations between [O~III] and H$\alpha$ images were presented in section 3.1.2, but the cross-correlation between the sharp filaments in the two lines found by the RHT provides a better measurement.  Figure~\ref{ccor_rht} shows the sum of the cross-correlations for all the rows in the [O~III] and H$\alpha$ RHT images.  It shows a clear peak at a lag of about 0\farcs5, extending to about 2\farcs0, which is qualitatively similar to the predictions of the 1D models.   In very rare cases, such as feature 1 in Figure~\ref{rht_diffo3ha}, the H$\alpha$ appears ahead of [O~III].  This is presumably a projection effect:  The filament is approximately a tangency of the shock to our line of sight, but there can be a ripple within that tangency, and different regions along the LOS can be bright in [O~III] and in H$\alpha$.  The difference in sharpness between the direct cross-correlation and the cross-correlation between the RHTs is presumably the result of diffuse emission in the lines.
 
  
  \begin{figure}
      \centering
      \includegraphics[width=0.7\hsize]{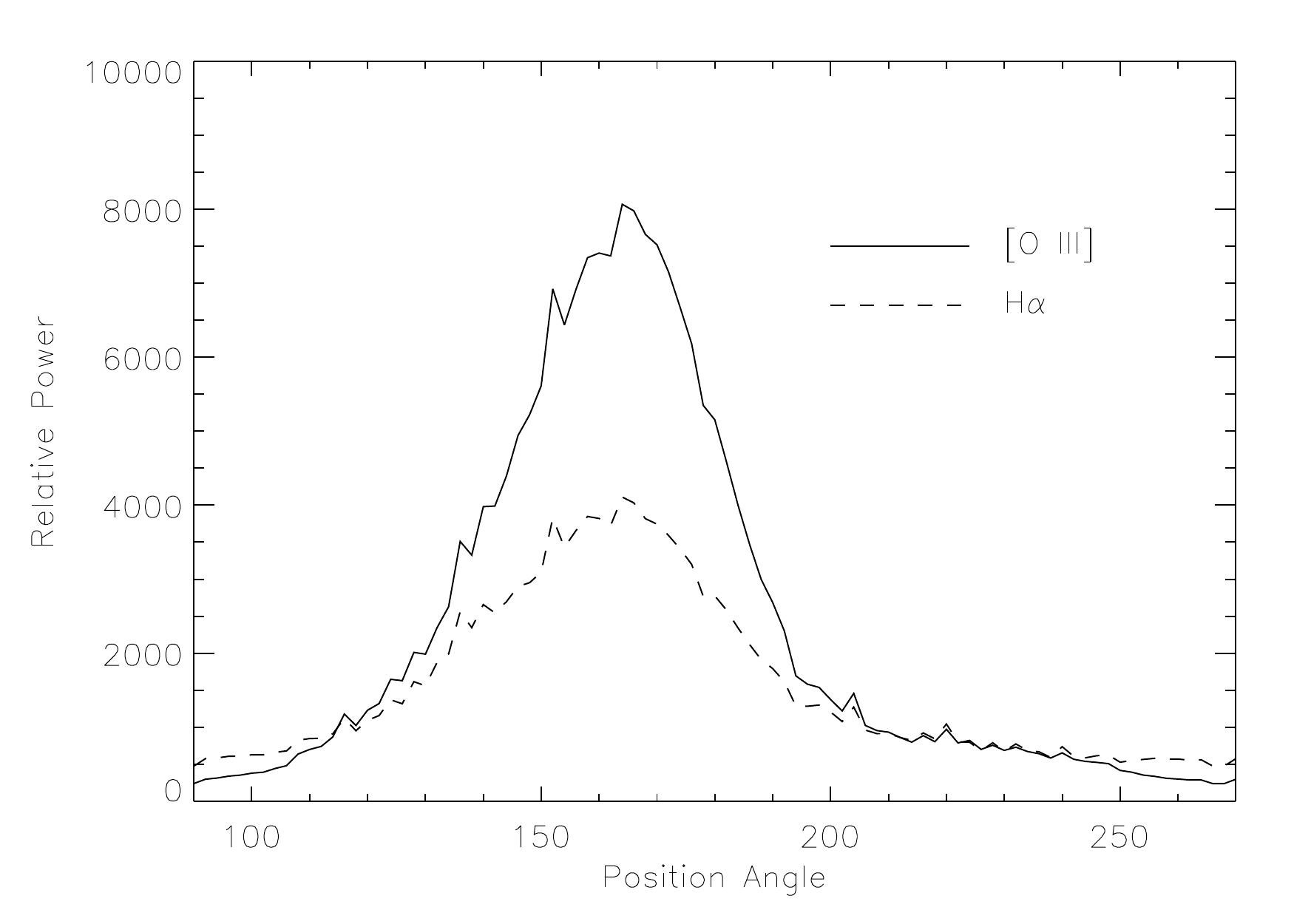}
      \caption{Relative power of the RHT as a function of angle for H$\alpha$ and [O~III] in arbitrary units.  Note that the RHT angles have a 180 degree degeneracy, and we have wrapped the plot around 180$^\circ$.}
      \label{sumthets}
  \end{figure}

  \subsubsection{Ripple amplitudes from RHTs}
  
 Figure~\ref{sumthets} shows the relative power in the RHTs of the [O~III] and H$\alpha$ images shown in Figure~\ref{rht_o3ha} as a function of Position Angle, summed over the images.  The units are arbitrary, but the comparison shows that the structure is aligned along a Position Angle of about 162$^\circ$.  That is within 1$^\circ$ from perpendicular to the Position Angle of the STIS slit, which was chosen to lie along the shock motion.  The FWHM of the H$\alpha$ angle distribution (35$^\circ$) is wider than that of the [O~III] distribution (30$^\circ$).  Most notably, the ratio of power at the peak to that perpendicular to the peak (filaments aligned along instead of perpendicular to the shock flow) is twice as large for [O~III] as for H$\alpha$.  This confirms the visual impression from the images that the [O~III] filaments are mostly aligned perpendicular to the shock direction, while many H$\alpha$ filaments are aligned nearly at right angles to it.
  
\begin{figure}
     \centering
     \includegraphics[width=3.4in]{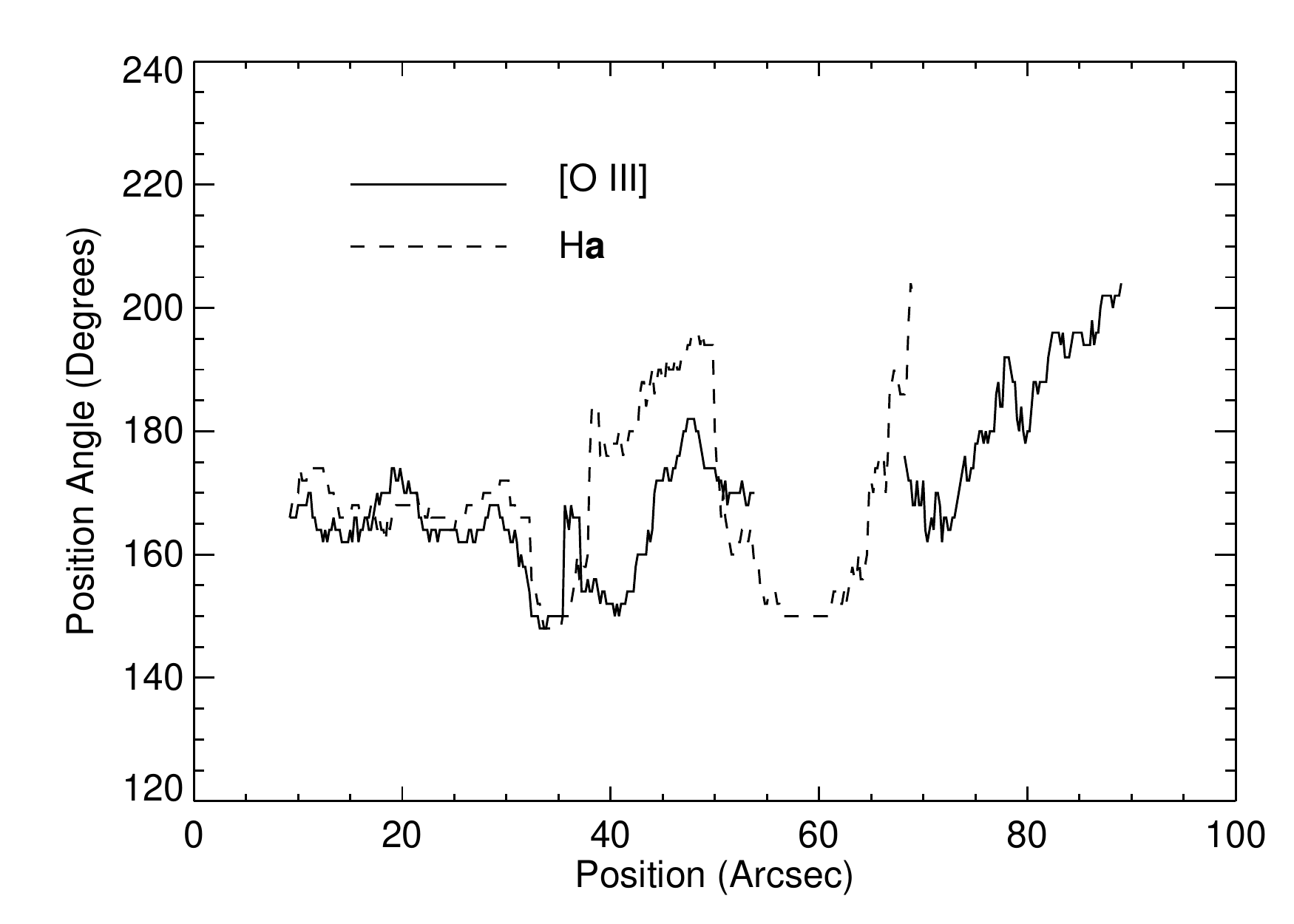}
     \caption{Position angle as a function of Y-position along the filament in the lower center from the [O~III] and H$\alpha$ RHTs (label 2 in Figure~\ref{rht_diffo3ha}).  The data have been wrapped around 180$^\circ$ because of the 180 degree degeneracy in the RHT.}
     \label{rht_angle}
 \end{figure}

 To quantify the amplitude of ripples, we can trace a filament and examine the local variations in angle given by the RHT.  At each position in the image, the RHT provides an RHT amplitude as a function of position angle at 2$^\circ$ intervals.  We follow a filament by stepping in the Y direction and choosing the X position of the peak RHT amplitude within a 15 pixel band centered on the current position.  For the RHT parameters we have chosen, the RHT amplitude as a function of angle tends to plateau at a maximum value, and we choose the center of this plateau as the local position angle.  Figure~\ref{rht_angle} shows the position angle as a function of Y-position along the filament indicated as '2' in Figure~\ref{rht_diffo3ha}.  Between about 55\arcsec\/ and 65\arcsec\/ there are two filaments, and the position angle in [O~III] is poorly defined.  The sharp jump at 36\arcsec\/ is a transition to a different ripple.  Overall, there are 10$^\circ$\/ variations on scales of 5\arcsec\/ and 30$^\circ$ variations on larger scales. The [O~III] filament in the north-central region shows smaller variations, with a total range of 20$^\circ$\/ and 10$^\circ$\/ to 15$^\circ$ variations on 10\arcsec\/ scales.
 
 A second way to quantify the amplitude of the ripples, is to examine the ``eye-shaped" features where both positive and negative excursions from the mean filament direction are seen.  Small, medium, and large examples are indicated with labels 3, 4 and 5 in Figure~\ref{rht_diffo3ha}.  The lengths and widths of 15 of these features with lengths ranging from 2\farcs6 to 26\farcs0 were measured by eye, and all but one fell in the range width/length = 0.18 to 0.30, which means a ratio of amplitude/wavelength=0.04 to 0.08.  Even the straight feature 'f' in Figure~\ref{ion-3panel} shows smaller features with similar amplitude/wavelength ratios.  That suggests that angles up to 30$^\circ$ from the average might appear.  This is clearly a subjective estimate, but it agrees with the overall width of the distribution of angles shown in Figure ~\ref{sumthets} and the estimate from the local variations in angle.
  
 \subsubsection{Velocities from RHTs}
  
Velocities can also be measured by determining proper motions from the RHTs of the 1997 and 2015 images.  As with the cross-correlations above, the RHTs give sharper peaks than the images themselves.  Unlike the images in Figures~\ref{rht_o3ha} and ~\ref{rht_diffo3ha}, we use RHTs at the full image resolution.  At each pixel above a chosen threshold in the 1997 RHT, we select the strongest RHT within $\pm$10 pix in x to avoid duplicating the measurement for the 5 or 10 pixels across the filament.  We make 80x80 subimages centered on the pixel and rotate them according to the RHT position angle and measure the position difference between the 1997 and 2015 peaks to get proper motion perpendicular to the filament.  As in Paper I, we assume a distance of 735$\pm$25 pc \citep{fesen18}, and we do not attempt to centroid the positions to subpixel scales because the 1997 images have already been resampled from 0\farcs1 to 0\farcs04.  At that pixel scale, the velocities are quantized in units of 7.93 \kms .

\begin{figure*}
  \centering
  \hspace*{-0.5in}
  \includegraphics[width=6.8in]{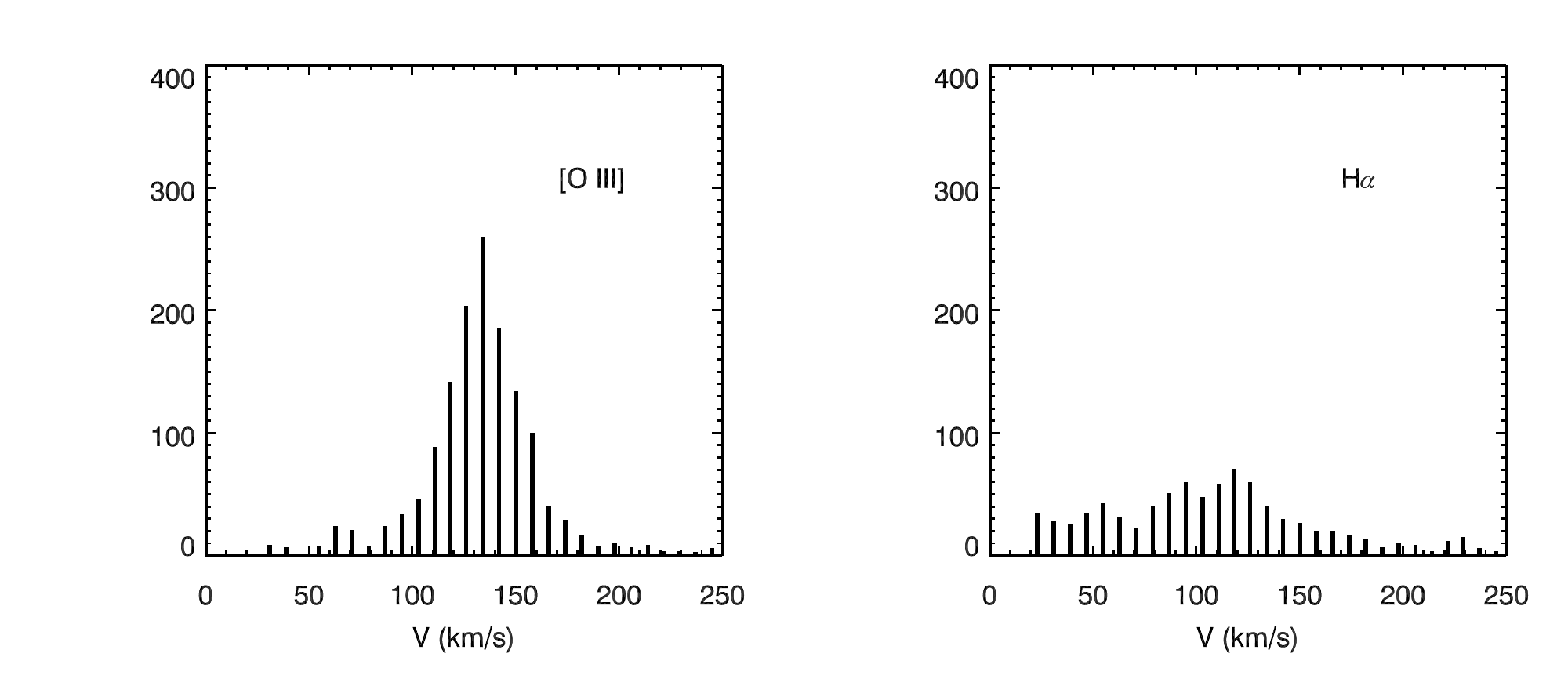}
  \caption{Histograms of velocities obtained by measuring proper motions from RHTs of 1997 and 2015 images in [O III] and H$\alpha$.  A distance of 735 pc was assumed.  These are the velocities perpendicular to the filaments.}
  \label{RHT_velocities}
  \end{figure*}

Figure~\ref{RHT_velocities} shows histograms of the [O~III] and H$\alpha$ velocities measured in the northern section of the WFPC2 images, covering both the eastern and western bright filaments.  The [O~III] velocities are smoothly and symmetrically distributed around the average speed of 130 \kms\/ found in Paper I.  The H$\alpha$ speeds are smaller and more broadly distributed.  A possible interpretation of the H$\alpha$ data is a broad peak centered at $\sim$110 \kms\/ plus a number of very slow filaments in the 25-70 \kms\/ range. 

Note that our procedure measures the proper motion velocity perpendicular to the local ($\pm$ 0\farcs8) filament, $V_\bot$.  A shock driven by gas pressure into a uniform medium will indeed move perpendicular to its surface, but density gradients along the shock front lead to oblique shocks.  For many purposes, $V_\bot$ is the relevant quantity, since it is the effective shock speed that determines the postshock temperature, the ram pressure, and other quantities related to the energy.  

To get a sense for whether the shocks are oblique, we can blink the images from different epochs to see whether the shock seems to be advancing perpendicular to its surface or advancing along its long dimension.  For the long, well-defined [O~III] filaments, the motion between the two epochs does not provide a definitive answer because the angles are modest and the displacement is only on the order of 2\arcsec , but overall, they seem to be moving perpendicular to their lengths.  However, some H$\alpha$ features clearly show oblique motions.  

\begin{figure*}
  \centering
  \includegraphics[height=2.5in]{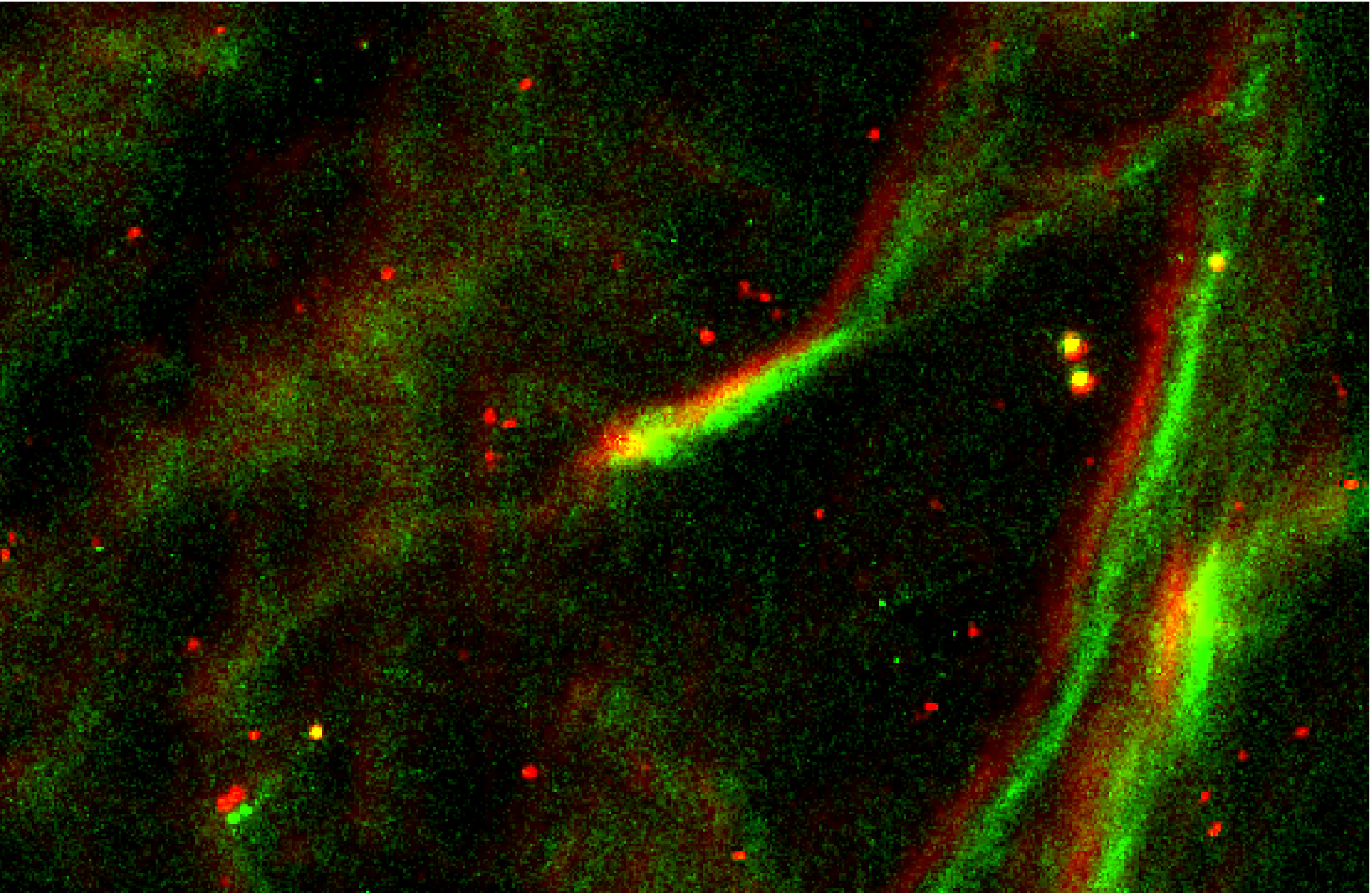}
*  \vspace*{1.2in}
\hspace*{-1em}
  \includegraphics[height=2.5in]{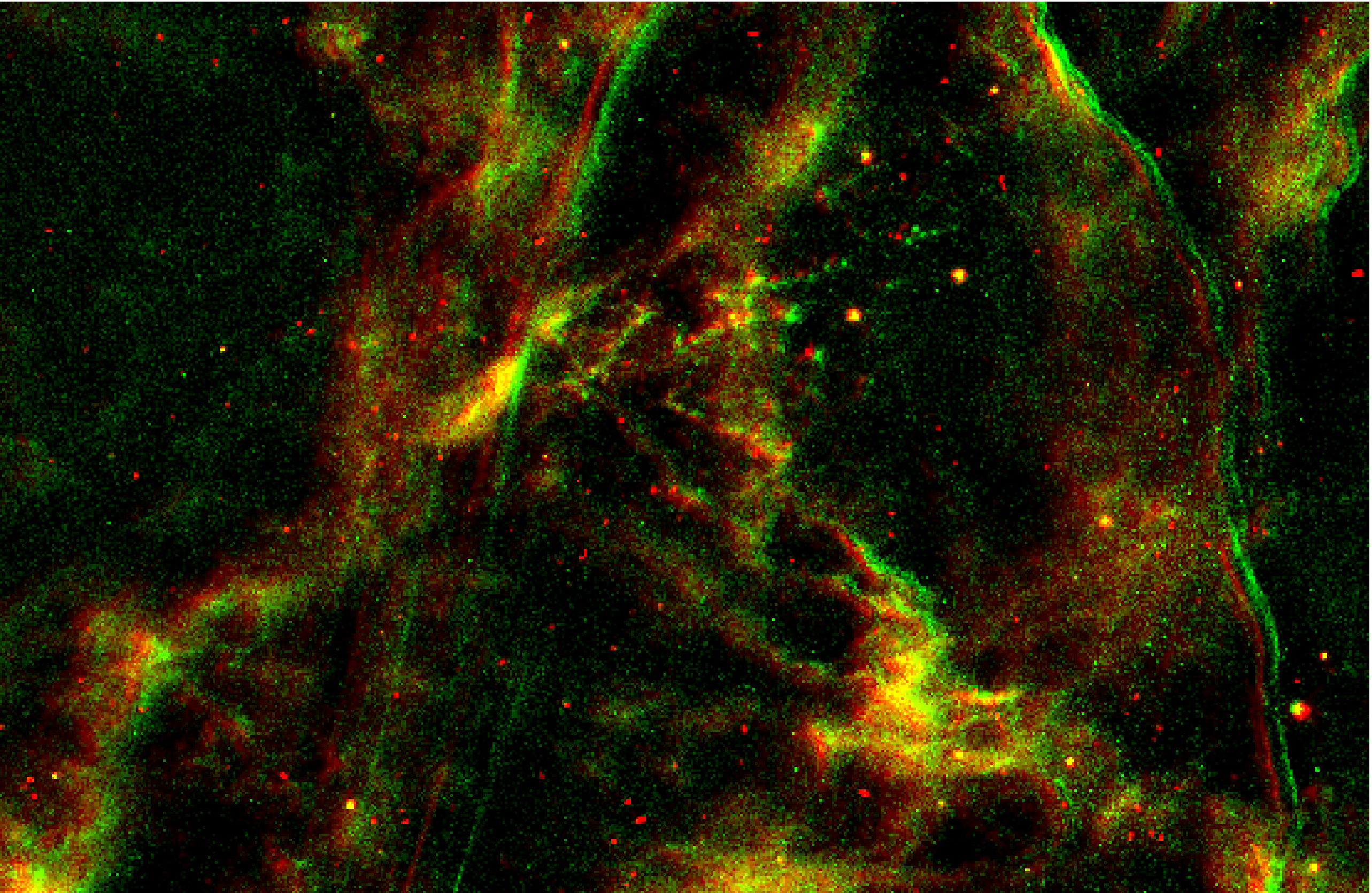}
 \vspace*{-0.8in}
  \caption{Overlays of the H$\alpha$ images from 1997 (red) and 2018 (green)images.  The first panel shows the feature 'a' of Figure~\ref{ion-3panel}, and the small knots north of center in the second panel are feature 'd' of Figure~\ref{ion-3panel}.  Yellow appears where the emission is bright in both epochs.}
  \label{ha_oblique}
  \end{figure*}

Figure~\ref{ha_oblique} shows overlays of the H$\alpha$ images from 1997 (in red) and 2018 (in green).  The northern part of the curved filament in the first panel (feature 'a' in Figure~\ref{ion-3panel}) appears to be moving perpendicular to its length, but the the southern part and the tip are moving East to West, nearly along the length of the filament.  The second panel shows a number of small, nearly pointlike knots (feature 'd' in Figure~\ref{ion-3panel})  That are also moving nearly East to West, though they lie along a feature elongated close to that direction.  This panel also shows a set of elongated, fuzzy features to the South of the knots, and those features also move obliquely when the images are blinked, with the perpendicular velocity component measured by our procedure about half as large as the parallel component.   

Both the knots and the larger fuzzy structures in the second panel of Figure~\ref{ha_oblique} may not be currently active shocks, but clumps that were shocked in the past and are now photoionized by radiation from other parts of the shock front.  They show no [O III] emission and [O I]:H$\alpha$ ratios above 1.0, indicating that the gas has mostly recombined.  According to the models, they are cooler than about 7,000 K, but they do not stand out in the temperature map of Figure~\ref{T_o2ha}.  Therefore, at least some of the lower proper motion velocity seen in H$\alpha$ can be attributed to our technique of measuring proper motions perpendicular to the filaments, while some H$\alpha$ features move obliquely. 

The proper motion velocities can also be used to determine the shock speed variations along a filament.  Figure~\ref{v_variation} shows the velocities measured along a section of the northeastern [O~III] filament, where there are relatively few filament crossings to complicate the measurement.  Even so, there are some outliers above 200 \kms\/ and below 100 \kms\/ that seem to be artifacts due to changing brightness substructure in the filament, and there are a number of points where no velocity could be determined.  This section shows a unique small (about 1\arcsec) region where the shock speed reaches about 180 \kms .

The shock speeds shown in Figure~\ref{v_variation} reveal gradients as high as 30 \kms\/ over scales of $10^{16}~\rm cm$, but more typically 15 \kms\/ over that scale, or $1.5 \times 10^{-10}~\rm s^{-1}$.  
Similar plots for H$\alpha$ or for the western [O III] filaments show similar gradients, but they are more difficult to analyze because multiple ripples cross each other (in projection), and because the complex structure causes more artifacts when the algorithm misidentifies pairs of filaments in the 1997 and 2015 exposures.  The shock speeds in the western filament vary more widely, from about 60 to almost 200 \kms .  As was found from the proper motions measured in Paper I, the velocities in the southern filament are around 30 \kms\ smaller than in the north. 

The velocity variations provide an important quantity.  The ratio of the velocity variance to the shock speed implies that the ratio of turbulent kinetic energy to the thermal energy of the shocked gas is about 5\%.  The plasma $\beta$ is large in the gas just behind the shock, so the contribution of magnetic energy to the energy content of turbulence is small.  The turbulent energy is a fundamental parameter for studies of magnetic field amplification and particle acceleration as the energy cascades to small scales \citep{yokoyama20, kamijima20}.  It is also important for estimating the degree of polarization of radio synchrotron emission, because strong turbulent ampification would imply a disordered magnetic field.  Strong small-scale tubulence can also mix cloud and intercloud material to enhance charge transfer emission in the X-rays \citep{lallement04}.  

\subsubsection{Vorticity from RHT velocities}

\begin{figure}
    \centering
    \includegraphics[width=4.4in]{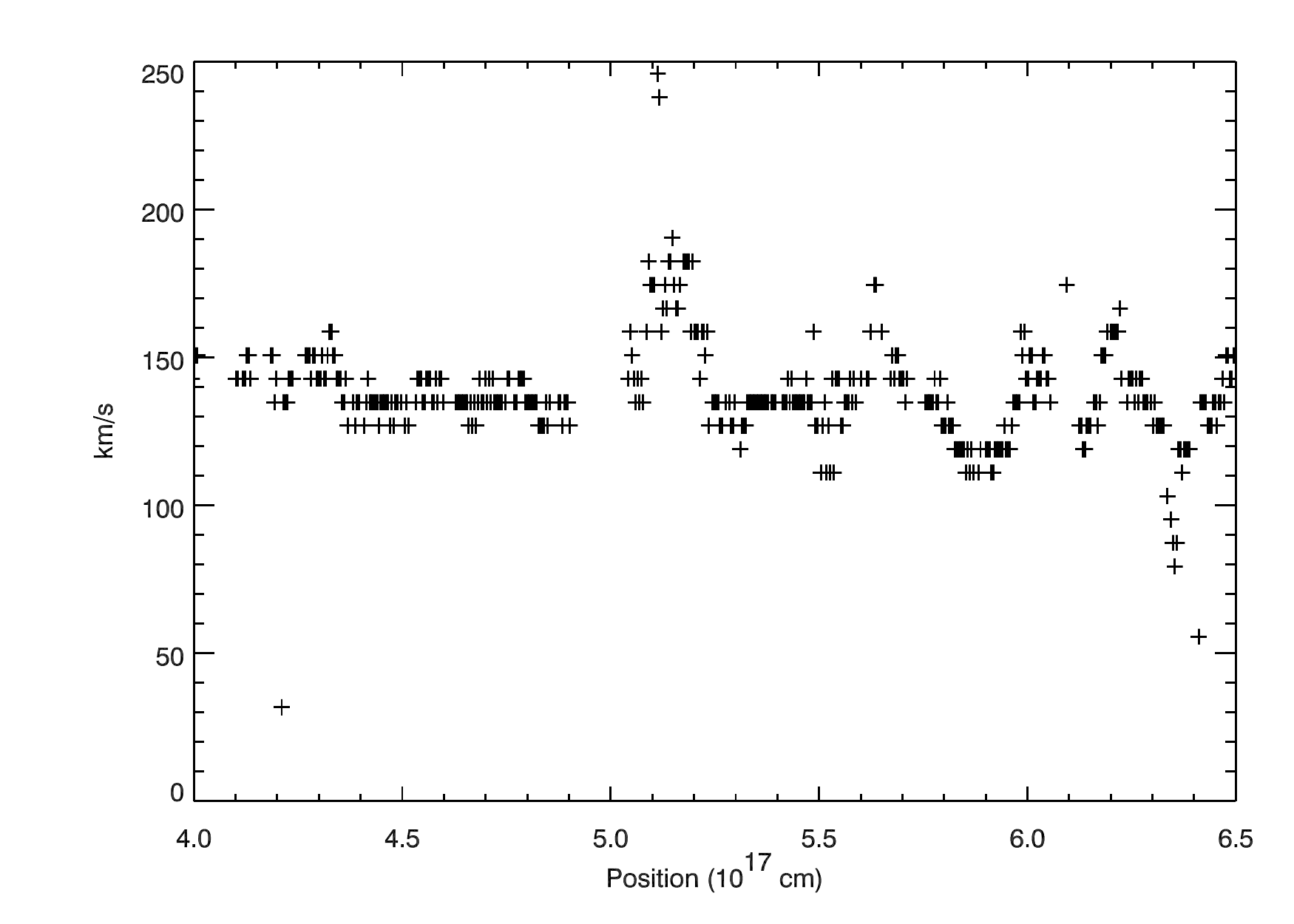}
    \caption{Proper motion velocities along a section of the northeastern [O~III] filament.  Outliers above 200 \kms\/ and below 100 \kms\/ are probably artifacts.}
    \label{v_variation}
\end{figure}

In Paper I we measured gradients of the Doppler velocity along and perpendicular to the Binospec slit ($dV_z/dy$ and $dV_z/dx$ in the coordinate system used there). We then used a symmetry argument that they equal $dV_x/dz$ and $dV_y/dz$ in order to estimate the vorticity component about the $y$ (shock motion) direction to be $2 \times 10^{-10}~\rm s^{-1}$.  That symmetry argument does not apply to terms involving $V_y$ because $y$ is the preferred direction, and we resorted to a less robust estimate of the gradients in $V_y$ to determine the components of vorticity perpendicular to the shock motion.  The gradients in Figure~\ref{v_variation} are $dV_y/dx$ in the coordinate system of Paper I, and by symmetry about the preferred $y$ direction, that equals $dV_y/dz$ on average.  Assuming that the gradients in the different directions are uncorrelated, the vorticity is $\sqrt 2$ times the measured gradient, so the vorticity is $2 \times 10^{-10}~\rm s^{-1}$ for eddies about a vector perpendicular to the shock direction.  That is close to the estimate in Paper I, but more directly based on the measurements.

The improved vorticity estimate provides an important parameter for the development of turbulence behind the shock, since it is the inverse of the eddy turnover time.  The resulting turnover time of about 150 years is comparable to the time it takes for the gas to cool to 10,000 K.  \citet{hrh87} give a cooling time of 6.8 years for a 130 \kms\/ shock with $n_0$ = 100, and the cooling time scales as 1/$n_0$, so $n_0$=6 would give a cooling time of 110 years.  The time to cool to the temperatures where [O~III] is produced is somewhat smaller, around 90 years.  The time it takes to cool from the [O~III] emitting region to temperatures around 7000 K where H$\alpha$ is produced is also around 90 years.  Thus, it is plausible that vorticity at the shock can account for much of the rippling of the [O~III] filaments and the more irregular appearance of the H$\alpha$ structure.

\subsection{Comparison of [Ne~IV] and [O~III]}

\begin{figure}
  \center
\includegraphics[width=0.65\hsize,trim={0cm 0cm 0cm 0},clip]{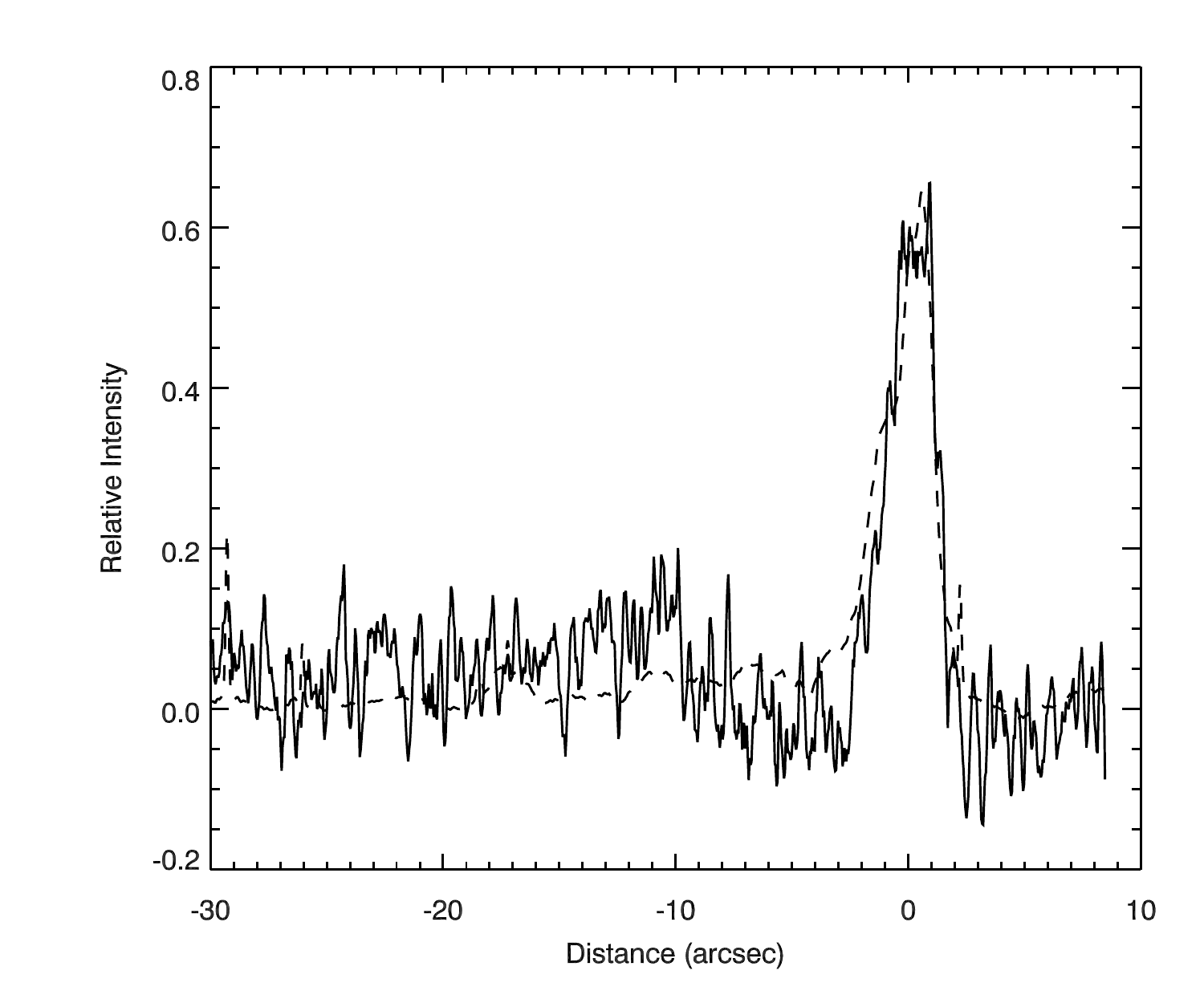}
\caption{Comparison of the intensities of [Ne~IV] $\lambda$2420 and [O~III] in a strip along the flow direction.  The solid line is [Ne~IV] and the dashed line is [O~III] multiplied by 0.02 in order to appear on the same scale.  The shock is moving toward the right.
\label{ne4_o3}
}
\end{figure}


The [Ne~IV] doublet is formed at $1.25\times 10^5$ K in ionization equilibrium, while the [O~III] line is formed at $8 \times 10^4$ K.  If turbulence develops as the gas cools, there should be some morphological difference between the [Ne~IV] and [O~III] images.  However, the low signal-to-noise ratio of the [Ne~IV] images precludes a real comparison. Instead, we sum the [Ne~IV] and [O~III] intensities over a 12\farcs5 slice perpendicular to the flow direction and plot the intensities as function of position along the flow direction in Figure~\ref{ne4_o3}.  We expect some spatial offset in the cooling flow behind the shock because the [Ne~IV] forms at a higher temperature,  and the 1D models in Paper I predict a $2.6 \times 10^{15}$ cm offset, or 0\farcs24.  While that is several times the WFC3 resolution, the $\sim$3\arcsec\ thickness of the main filament in these lines and the low S/N in the [Ne~IV] data mean that it cannot be reliably measured. However, an offset of this magnitude is consistent with the data.

\subsection{Spectra}

Plots of the FUV and NUV spectra were shown in Figure~\ref{uvspec}, and the intensities measured from a 3\farcs7 wide section of the slit are presented in Table 3.  As can be seen in Figure~\ref{uvspec}, the UV lines are very wide because we used the 2\arcsec\ slit to maximize the count rates.  That blends the He~II $\lambda$1640 line with the O~III] $\lambda$ $\lambda$1661, 1666 doublet.  As discussed above, the G140L spectrum shows two comparable flat regions on either side of 1630\AA\/ plus a region in the center that is equal to the sum of the two.  The 3\farcs7 wide band that was measured covers the two bright features near the center of the STIS slit shown in Figures~\ref{o1_o2} and \ref{o3_ha}.  

\subsubsection{Comparison with a shock model}

The shock model shown in Table 3 for comparison to the spectra was calculated with the updated version of the code used by \citet{raymond79}, and it is similar to those tabulated in Paper I, except that it is more severely truncated to simulate an incomplete cooling region \citep{raymond88}.  It again assumes a shock speed of 130 \kms , pre-shock density of 6.0 $\rm cm^{-3}$ and magnetic field of 4 $\mu$G. but the integration is cut off when the gas cools to 8300 K, rather than 6700 K as in Paper I. The more severe cutoff is required by the very high ratios of [O~III] and the UV lines to the Balmer lines.  Such severe incompleteness is not surprising in that we have selected a very bright [O~III] region with little emission in the cool lines.  

Table 3 shows that many of the observed lines agree well with the model.  A test of the scaling between optical and UV lines is the O~III] to [O~III] ratio (1664\AA\/ and 5007\AA ), which depends only weakly on model parameters.  The model agrees with the data to within 30\%, which is easily within the uncertainty in the reddening correction for two lines widely separated in wavelength.

The apparent overprediction of C~IV results from resonance scattering in the nearly edge-on sheet of gas \citep{cornett92, raymond03}.  That process also affects the Si~IV doublet at 1400 \AA\/, but Si~IV only contributes about 10\% of the blend with O~IV].  Depletion of the refractory elements C and Si onto grains could be a factor, but the C~III] to O~III] ratio indicates that carbon is depleted by no more than 25\% .  That is in general agreement with the result of Paper I that a significant fraction of the iron is returned to the gas phase when the gas is still hot enough to produce the [Fe~VI] and [Fe~VII] optical lines.  The He~II $\lambda$1640 line is underpredicted by the model.  That could indicate that some of the He in the preshock gas is singly ionized, while the model assumed that is it doubly ionized.  When singly ionized He is excited in the thin ionization zone just behind the shock, it increases the $\lambda$ 1640 intensity \citep{coxraymond85}.  The preshock ionization state is difficult to predict in a case where the shock speed is changing rapidly because of the encounter with a cloud of unknown density profile.

There are other discrepancies between observations and models in Table 3.  The [O~II] $\lambda$3727 doublet and the C~II]/Si~II] blend at 2325 \AA\/ are overpredicted, while the [O~I] line is underpredicted.  The peak in the reddening correction in the 2200 \AA\/ would affect the C~II]/Si~II] feature, so uncertainty in the reddening might be part of the problem.  However, the main difficulty is that simply cutting off the model when the gas cools to 8300 K is a badly oversimplified way to treat the incomplete cooling and recombination zone, given the complexity of the optical images and the projection of different emission regions along the line of sight.  A higher-temperature cutoff would help the [N~II], [O~II] lines and the C~II]/Si~II] blend, but it would worsen the underprediction of the [O~I] line.  A similar, though less severe, difficulty arose in Paper I.  It is likely that the high and low ionization lines come from unrelated regions that are seen together in projection.  It would be possible to add together sets of models with different shock speeds and low temperature cutoffs to match the observations, but the results would probably not be unique or useful.

\subsubsection{Ram pressure}

Paper I used the method of \citet{raymond88} to determine the ram pressure.  Radiative shocks in the 100 to 150 \kms\ range produce 0.55$\pm$ 0.05 [O~III] $\lambda$5007 photons for each H atom that passes through the shock.  The surface brightness of a face-on shock is therefore  0.55 $\rm n_0 V_s / 4 \pi$ photons/($\rm cm^2~s~sr$).  If the shock is not face-on, its brightness is increased by a factor $\rm 1 / cos\theta$.  The observed Doppler velocity is $\rm V_s cos\theta$, so $\rm cos \theta$ cancels out in the product $\rm I_{obs} V_{obs} = 0.55 n_0 V_s^2/4 \pi$, which is a constant times the ram pressure, $\rho V_s^2$.  This method cannot be used on the bright filaments, because those are places where the LOS is tangent to the rippled shock front \citep{hester87}, so the LOS component of velocity is zero.  It can be used in the fainter regions between the bright filaments where the shock velocity has a component along the LOS.

\begin{figure}
  \center
\includegraphics[width=0.4\hsize]{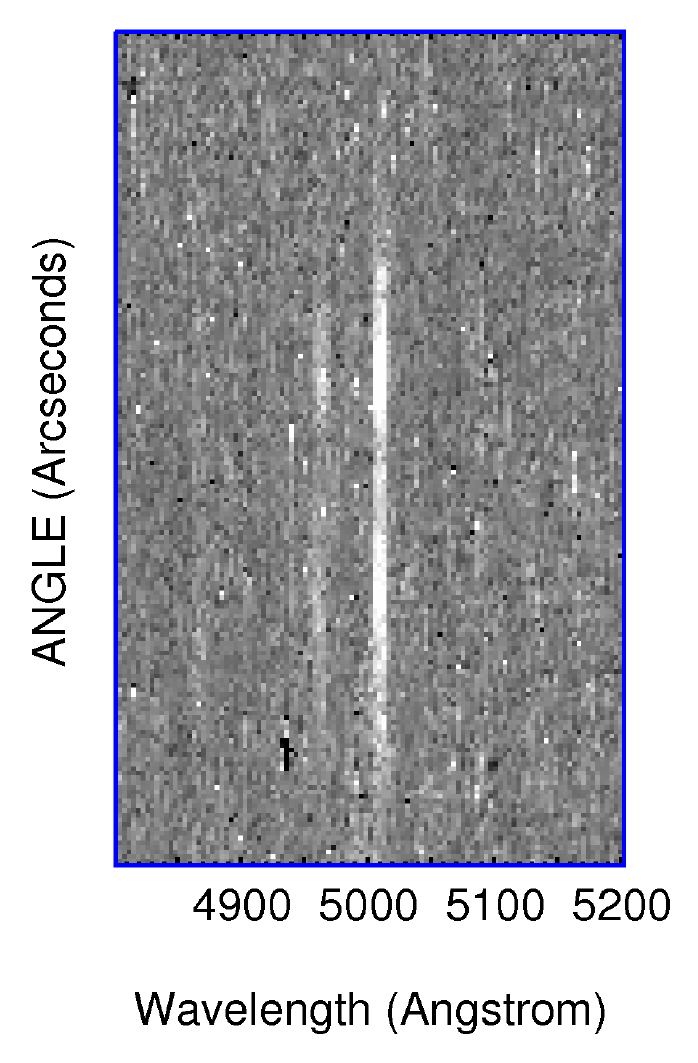}
\caption{The section of the G430M spectrum with the 0\farcs1 slit showing the $\lambda \lambda$ 4959, 5007 [O~III] lines.  The vertical scale extends from -8.23 to +3.96 relative to the reference position.
\label{oiii_G430M}
}
\end{figure}

A limitation of that method is that the LOS may pass through a rippled sheet more than once, and positive and negative velocities can cancel,  leading to an underestimate of the ram pressure.  The high spatial resolution of HST provides a partial check on this possibility.  A G430M spectrum was obtained with a 0\farcs1 wide slit in order to measure the intensity and Doppler shift of a region in between the two bright regions along the slit.  Figure~\ref{oiii_G430M} shows the [O III] portion of the STIS spectrum. The faint region between about -3.0\arcsec\ and -1.9\arcsec\ is slightly redshifted compared to the brighter regions above and below it, which are close to tangency to the LOS. The measured velocity shift of 29 \kms\ and surface brightness of $2.9 \times 10^{-15}~ \rm erg~cm^{-2}~s^{-1}~arcsec^{-2}$ (including extinction correction) imply a ram pressure of $3.1 \times 10^{-9}~\rm dyne ~cm^{-2}$, 22\% smaller than in paper I, and therefore a smaller preshock density of 4.7 $\rm cm^{-3}$. This also gives a slightly smaller preshock magnetic field (perpendicular component) of about 5 $\mu$G.  

This measurement pertains to a different region than the slit position used in Paper I, which was about 15\arcsec\/ farther north.  There is no apparent blueshifted contamination in the STIS spectrum that might reduce the apparent ram pressure, and ram pressure variations at least as large are 25\% are expected within the filament complex.

\section{Discussion}  

We have set out to understand the development of turbulence in the radiative shocks in the Cygnus Loop, in particular the different morphologies of the H$\alpha$ and [O~III] optical lines.  In this section we discuss four mechanisms in the light of the shock parameters from Paper I and the images and spectra analyzed above.  We consider the roles of thermal instability, the thin shell instability, inhomogeneities in the preshock density and variations in the preshock magnetic field.

\subsection{Thermal Instabilities}

\citet{field65} discussed thermal instabilities in the solar atmosphere and interstellar gas, and the idea was applied to the cooling regions behind shocks by \citet{innes87} and \citet{gaetz88}.  For a constant-density medium without a magnetic field, they found an oscillating shock speed.  A region of slightly higher density and lower temperature cools more quickly than the surrounding gas.  If the scale of the region is set by the cooling length, the sound crossing time can exceed the cooling time, and the cooling gas can fall out of pressure equilibrium with its surroundings.  As a result of constant-density cooling, the gas will release only 3/2 kT per particle instead of 5/2 kT for constant pressure cooling.  In addition, the low pressure cooled gas will be subject to secondary shock waves driven by the higher pressure surrounding gas.  The overall result is that low temperature lines such as [O~I] and [S~II] will be enhanced relative to higher temperature lines such as [O~III].

The detailed numerical simulations by \citet{innes92} and \citet{sutherland03} explored the effects on the emission line spectra and the velocity structure.  The instability sets in for shocks faster than about 120 - 150 \kms , because the cooling coefficient increases steeply with temperature for temperatures below roughly log T = 5 \citep{raymond76}.  It takes several cooling times for the instability to develop, so \citet{innes92} used a preshock density jump to initiate the process.  Once it has developed, the thermal instability produces factor of two changes the average optical line ratios and much larger variations at specific times.  The models by Innes show multiple intensity peaks along the flow direction.  However, it is difficult to distinguish multiple intensity peaks from projection effects within a rippled sheet seen nearly edge-on.

The thermal instability operates on the scale of the cooling length in the flow direction, but the secondary shocks can create smaller-scale structures in the cooler lines, such as H$\alpha$ \citep{innes92}.  The structures will have a similar scale in the transverse direction \citep{bertschinger86}.  The thermal instability will be cushioned by magnetic pressure, softening any secondary shocks \citep{innes92}, and magnetic pressure supports the gas in the photoionized region below about 10,000 K in this part of the Cygnus Loop (Paper I).

Overall, the shock speed is at the low end of the range where the thermal instability operates, meaning that it should not be very strong, and the incomplete cooling of much of the shocked gas implies that the thermal instability would barely have time to develop.  Moreover, the scale is expected to be on the few arcsecond scale of the cooling length, while much larger features are seen in the images.  However, it is possible that preshock density fluctuations trigger the thermal instability, and the scale of those fluctuations could determine the length scale of the observed features.

Thermal instability does offer a plausible explanation for one otherwise mysterious feature of the proper motion velocities.  As described above, the speeds derived from the proper motions in the [O~III] and H$\alpha$ images typically differ by 10 to 20 \kms .  In the north, the [O~III] is moving faster, while in the south the H$\alpha$ proper motions are higher.  That fits in with the prediction of thermal instability models that the shock speed oscillates about a mean value, and depending on the phase in the oscillation, the shock or the dense material driving it may be moving faster.  Also the secondary shocks predicted by the numerical simulations could explain the sharpness of many of the H$\alpha$ features compared to the diffuse $\sim$2\arcsec\/ wide H$\alpha$ emission predicted by the 1D models.

Keeping in mind that most of the measured velocities are at, or even below, the 120 \kms\ threshold for the thermal instability \citep{sutherland03}, the instability would be expected to develop slowly, if at all.  However, it is highly likely that the shock is slowing down rapidly because of loss of pressure in the driving gas \citep{straka}.  It is also likely that the shock is slowing as it enters a cloud and is therefore probably climbing up a density gradient, as was seen on the eastern limb of the Cygnus Loop \citep{szentgyorgyi00}.  The structure we see may therefore be partially a relic of thermal instabilities that developed in a faster shock in the recent past.  

We conclude that thermal instability does not play a dominant role in the early development of the rippled sheet morphology seen in [O~III], but that it may account for some of the complex structure seen in H$\alpha$, especially at small scales.

\subsection{Thin shell instability}

The thin shell instability \citep{vishniac83}, which is actually an overstability, arises when a dense shell is being pushed through the ISM by the pressure of hot gas.  As long as the thermal pressure force is anti-aligned with the shock ram pressure, the two can balance.  However, a ripple in the shell causes the pressure force (perpendicular to the shell locally) to be misaligned with the ram pressure (perpendicular to the shell globally), causing the ripple to grow.   The instability occurs on scale lengths larger than the shell scale height and grows most quickly on small scales.  The growth of the thin shell instability is inhibited by large-scale magnetic fields \citep{ryu91, riley20}.

In the case of the western Cygnus Loop, the dense shell is just beginning to form, and there has not been time for the thin shell instability to develop.  The high [O~III] to H$\alpha$ ratios generally indicate incomplete shocks that have existed for only a few cooling times.  Therefore, the shell is just beginning to form, and the thin shell instability is unlikely to play a dominant role.  Moreover, magnetic pressure dominates in the layer of cool postshock gas (Paper I), so the growth of the instability will be severely reduced if the field is coherent. 

We conclude that the thin shell instability does not play a major role in the development of the postshock structures in the western Cygnus Loop, though we have not been able to identify a clear signature of this instability that would provide a definitive test.  The basic physics of misalignment between the ram pressure and the internal thermal pressure may amplify the effects of the other instabilities, however. 


\subsection{Density inhomogeneities}

To first order, the ram pressure, $\rho_0 V_S^2$, is constant over the surface of an SNR blast wave, so the local shock speed varies as $n_0^{-1/2}$.  The reduced shock speed in low amplitude density enhancements will produce the rippled structure described by \citet{hester87} and apparent in the nonradiative Balmer line filaments in the northern Cygnus Loop \citep{salvesen09, medina14} and in the the [O~III] image discussed here.

Discrete clumps with high density contrast are generally called clouds, and their interaction with a blast wave has been simulated by \citet{klein94} and many subsequent works. Clear examples in the Cygnus Loop are the SE cloud \citep{fesen92}, a cloud to the south \citep{patnaude02}, and a cloud undergoing the transition from nonradiative to radiative in the northeast \citep{raymond83, long92, hester94, blair99, sankrit00}.  The high density contrast means that the shock in the cloud is severely decelerated, and the blast wave sweeps past the cloud.  Moderate or high density constrast fluctuations can generate strong shear and vorticity, creating small-scale turbulence, and winding up and amplifying the magnetic field \citep{guo12, fraschetti13}, accelerating particles \citep{fraschetti15, zank15}, and eventually shredding the cloud. 

We do not directly observe the preshock density inhomogeneities, but assuming that the ram pressure is roughly constant we can infer them from variations in the shock speed.  The proper motions of the north-central [O~III] filament vary from 120 to 180 \kms\/ along its 60\arcsec\/ length, with 40 \kms\/ variations on length scales as small as 5\arcsec\/ (Figure~\ref{v_variation}).  If the ram pressure is constant, the preshock density varies by up to $\pm$40\% on those scales, though $\pm$20\% is the typical value.  The northern half of the western filament shows a similar range, 120 to 142 \kms , while the southern half is generally slower, 95 to 125 \kms .  Filaments behind the main western filament are significantly slower, but they are seen in projection, and they may lie at significant distances from the main filament along the LOS direction.
The H$\alpha$ velocities show a similar range with a lower average speed, except that there are a number of very slow features in the $\sim$25 to 70 \kms\/ range (Figure~\ref{RHT_velocities}).  These slower shocks may be part of a slower shock front moving roughly south to north and seen in projection against the main faster shock.  It is also possible that they are relic shocks that have rapidly slowed down.  However, as discussed in section 3.3.3, at least some of the apparently low velocities are actually the perpendicular components of faster oblique motions.

Shock speed variations above 20\% would distort the shock much more severely than is seen in the [O~III] image if they persisted very long.  The ripple amplitude/wavelength ratios of 0.04 to 0.08 and angular variations of 10$^\circ$ to 20$^\circ$ suggest that the scales of the density variations are similar along and perpendicular to the shock direction.  That is not necessarily surprising, but a magnetically dominated density structure at the cloud boundary could well have different scales in different directions. 

Variations in preshock density larger than 20\% to 40\%  would produce larger variations in shock speed than observed.  We conclude that density variations are typically $\pm$20\%.  There is no evidence for discrete clouds with large density contrasts analogous to those mentioned above that are seen in other parts of the Cygnus Loop in the region we investigate here.  Emission in the SW corner of Figure~\ref{overlay} that is seen in the lower-ionization lines may come from a high density contrast cloud, but it is not included in our analysis.  It appears ahead of the main shock in projection, but it is not apparently connected to the cooling regions we are studying.

The modest amplitude of the density inhomogeneities inferred from the [O III] rippling is consistent with the result that the vorticity is comparable to the inverse cooling time.  Both suggest that density inhomogenities account for much of the [O~III] rippling and the difference between [O III] and H$\alpha$ morphologies but that velocity shear has not yet produced a turbulent cascade.

\subsection{Magnetic inhomogeneities}

The magnetic field in the preshock gas is presumably irregular, and while we estimated an average strength of 5 $\mu$G for the perpendicular component, the field structure is unknown.  Even if the spectrum of field fluctuations in the molecular cloud were known, it could be modified by the shock precursor \citep{bell04, wagner09}, the shock itself \citep{capriolispitkovsky14}, and the vorticity that develops as the shock traverses an irregular density field.

In any case, we expect that there will be regions where the field is nearly perpendicular to the shock normal and places where it is nearly parallel.  In the former, the field is compressed by a factor of 4 at the shock, while in the latter, it is not compressed at all.  There will also be places where the field strength happens to be close to zero. Just behind the shock, the B field does not matter very much, because the plasma beta is high ($\sim 68$ from Paper I).  However, when the gas cools and loses thermal pressure, the gas and field are compressed until magnetic pressure dominates. In the magnetically dominated region, the $\bf B$ field variations could cause local velocity variations up to roughly the Alfv\'{e}n speed, or about 30 \kms. 

In the magnetically dominated region, where H$\alpha$ is formed, the field must reach a constant pressure.  Therefore, regions where the field at the shock was parallel to the flow must be compressed 4 times more than regions where it was perpendicular, and regions where the field happened to be weak can be compressed even more.  These denser regions should show up in H$\alpha$ images and might account for the fluffy structure observed \citep{raymondcuriel}.  In particular, the persistence of H$\alpha$ emission far behind the [O~III] filaments is difficult to explain unless magnetically structured cool gas is exposed to ionizing radiation, or unless it arises from a separate set of slow shocks seen in projection near the [O~III] filaments.  In the former case, the ionizing radiation could be emission from upstream that penetrates through a low-density magnetically supported matrix, or it could be X-ray emission from the hotter gas nearby.

Most of the emission in the hotter lines seems filamentary, and most of the emission in the cooler lines seems either filamentary or diffuse.  Some examples of very small knots are best seen in H$\alpha$ as feature 'd' in Figure~\ref{ion-3panel} and ~\ref{ha_oblique}.  They appear to be bright spots along a few of the minor filaments, and they could be regions of low B field that are more highly compressed.  The majority of the bright knots in the [O~III] image are simply places where two or more filaments cross in projection.  

The most likely examples of high-density photoionized clumps are seen in  H$\alpha$ in the lower panel of Figure~\ref{ha_oblique}.  The small knots labeled feature 'd' in Figure~\ref{ion-3panel} move obliquely, and the somewhat fuzzy features to the south of feature 'd' are more extended along their direction of shock motion than perpendicular to it.  The proper motions perpendicular to the structures are fairly small, but blinking the images shows that the motions along the long dimensions of the features are roughly twice as large.  The [O~I] to H$\alpha$ ratio indicates that the gas has mostly recombined, and there is little corresponding [O~III] emission.  These are probably not active shock fronts, but regions that were shocked in the past and are now photoionized by more distant parts of the shock.  These diffuse clumps, whose shapes do not correspond to the shape of the shock front, are the types of structures expected when the shock encounters low magnetic field regions and compresses the gas more strongly than the surrounding regions. The differences between [O III] and H$\alpha$ velocities are also consistent with the motions approaching the lfv\'{e}n speed expected from inhomogeneities in the magnetic field.

We conclude that magnetic inhomogeneities in the upstream gas play a role in the downstream structure seen at the lowest temparatures, though they cannot account for the smooth [O III] structures or the filamentary H$\alpha$ morphology seen in much of the region.

\section{Summary}

While some old SNRs show simple, gently curved sharp filaments in the optical emission lines, the Cygnus Loop and some other SNRs show very complex structure, and the smooth, sinuous rippled sheet morphology seen in [O~III] gives over to a more chaotic, diffuse appearance in H$\alpha$ and other lines formed at lower temperatures.  We have studied a portion of the western Cygnus Loop in order to investigate the reasons for this difference and to understand the development of turbulence in the cooling flow behind a radiative shock wave.  An earlier paper concentrated on a set of ground-based spectra from the Binospec instrument on the MMT (Paper I), and in this paper we have concentrated on images and spectra from HST.

Paper I concluded that the shock in this region moves at 130 \kms\/ based on the proper motions and known distance to the Cygnus Loop.  The density of the preshock gas is 6 cm$^{-3}$, with a perpendicular magnetic field strength of 6 $\mu G$.  The ram pressure was found to be $4 \times 10^{-9}~\rm dyne~cm^{-2}$, and the vorticity was estimated to be $2 \times 10^{-10}~\rm s^{-1}$. The optical spectrum was in fairly good agreement with a 1D model of a 130 \kms\/ shock truncated to simulate an incomplete recombination zone.  We also found that simple adiabatic compression of the ambient cosmic-ray population, along with compression of the gas and magnetic field in the postshock flow, could account for the nonthermal radio synchrotron emission and the pion decay gamma rays. That suggests that neither diffusive shock acceleration nor turbulent amplification of the magnetic field play a strong role in this relatively slow shock.

In this paper, we find that the UV spectrum agrees quite well with the prediction of a 1D 130 \kms\/ shock model, suggesting that the 1D models can be used to infer elemental abundances and depletions onto grains.  This is somewhat surprising given the complexity of the region and the very different morphologies in different spectral lines.  However, the use of a single 1D shock model with a low-temperature cutoff to account for an incomplete recombination zone is a fairly crude approximation, and it leads to factor of 2 discrepancies in the ratios of some lines of neutral and singly ionized species to the Balmer lines. The 1D models probably do better for the much simpler shock structures seen in some other SNRs.  We have also used the higher spatial resolution of the HST optical spectrum to obtain a new measurement of the ram pressure, which is slightly below the value given in Paper I, but is easily within the likely range of variation expected along the shock. 

To investigate the morphology, we have quantified the structure in several ways.  We computed cross-correlations, Fourier power spectra, and, most notably, Rolling Hough Transforms \citep{clark14}.  We find that the picture of the shock as a rippled sheet seen nearly edge-on \citep{hester87} accounts very well for the [O~III] filaments.  From the amplitude of the ripples in position and angle, we conclude that they are caused by density inhomogeneities of order $\pm$20\% on scales between (at least) $10^{16}$ and $10^{17}$ cm.  That picture also accounts for the variations in shock speed derived from proper motions obtained from the RHTs.  

Many of the [O~III] filaments have corresponding H$\alpha$ filaments trailing by 0\farcs5 to 2\farcs0, as predicted by the 1D models, but many do not.  The proper motions of the [O~III] filaments indicate shock speeds around 130 \kms , as was found in Paper I.  Where corresponding filaments are seen in the two lines, their proper motions indicate speeds that differ by 10 to 20 \kms .  Many of the H$\alpha$ filaments without [O~III] emission show much lower proper motion speeds, but at least some of them are actually faster clumps of emission moving obliquely.  Features seen in [O III] or H$\alpha$, but not both, could result from thermal instabilities or from the shock running through a small density enhancement.   

The variation in shock speed along each filament provides a velocity gradient component needed to make a more robust estimate of the vorticity ($\nabla \times$V), which is useful for determining the roles of various processes in the growth of turbulence in the cooling flow.  We find a value close to that given in Paper I, corresponding to an eddy turnover time scale of about 150 years.  Comparison with the cooling times indicates that vorticity at the shock could account for much of the rippled structure seen in [O~III] and the more irregular structure seen in H$\alpha$.  

The probable source of the measured vorticity is preshock density inhomogeneities of $\sim$20\% amplitude.  The [O~III] ripples show similar ratios of amplitude to wavelength over the observed range of scales.  This suggests that the upstream density variations do not follow a Kolmogorov spectrum.  It also suggests that a turbulent cascade to small scales does not develop within the cooling time of the gas, and similarity of the turnovers in the autocorrelation functions of [O~III] and H$\alpha$ at 1\arcsec\/ to 2\arcsec\/ (Figure~\ref{acorr}) supports that idea.  The lack of a Kolmogorov turbulent cascade could result from insufficient time, since there is $\lesssim$ one eddy turnover time available.  However, it is also quite likely that the large-scale magnetic field has begun to dominate the dynamics by the time the gas reaches the recombination zone, and the field suppresses the development of turbulence.  As in Paper I, we find that turbulent winding of the magnetic field and turbulent particle acceleration are not required to account for the inferred magnetic field strength or the observed nonthermal radio and gamma-ray emission.

We find that density inhomogeneities in the upstream gas dominate the formation of the rippled structure seen in [O~III].  The measured vorticity is enough to account for some structural change between the [O~III] and H$\alpha$ emitting regions, but not all.  Thermal instabilities and variations in the upstream magnetic field have little effect on the [O~III] structure, but they probably contribute to the structure seen in H$\alpha$.
Magnetic field inhomogeneities probably contribute to some of the small H$\alpha$ structures, but not to the large-scale filamentary structure.  We do not expect that the thin shell instability has much effect on the structure because the shell is still forming in the region we observe, but we have not identified clear signatures of that process to confirm or rule out its presence.








\section{ACKNOWLEDGEMENTS}

We thank Dr. Susan Clark for helpful comments on the manuscript.  This work was supported by HST Guest Observer grant GO-15285.0001\_A to the Smithsonian Astrophysical Observatory.  I.C. is supported by the Telescope Data Center at SAO and also acknowledges the Russian Science Foundation grant 19-12-00281 and the Program of development of M.V. Lomonosov Moscow State University for the Leading Scientific School ``Physics of stars, relativistic objects and galaxies.''. B.B. acknowledges the generous support of the Simons Foundation. We obtained HST/WFPC2 images from the MAST archive at STScI. CHIANTI is a collaborative project involving George Mason University, the University of Michigan (USA), University of Cambridge (UK) and NASA Goddard Space Flight Center (USA).

%

\vspace{5mm}
\facilities{HST(STIS), HST(WFC3), HST(WFPC2), MMT(BINOSPEC)}


\software{RHT \citep{clark14}, SWarp \citep{bertin02}}




\bibliography{main}{}
\bibliographystyle{aasjournal}



\end{document}